\documentclass[aps,prb,nopacs,floatfix,12pt]{revtex4}

\usepackage{graphicx}
\usepackage{amssymb}
\usepackage{amsmath}
\usepackage{epsfig}
\usepackage{bm}
\usepackage{mathtools}
\graphicspath{{/home/sgreben/FIG/CO2.SLAC/}}

\begin{document}
\bibliographystyle{jcp}

\title{Quantum mechanical study of the attosecond \\
  nonlinear Fourier transform spectroscopy of carbon dioxide}

\author{Sergy\ Yu.\ Grebenshchikov\footnote{Email: sgreben@gwdg.de}}
\affiliation{R\"usterstr. 24, 60325 Frankfurt am Main, Germany} 
\author{Sergio Carbajo}
\affiliation{SLAC National Accelerator Laboratory and Stanford University,
  2575 Sand Hill Rd, Menlo Park, CA 94025, USA}

\begin{abstract}

  \noindent
Attosecond 
  nonlinear Fourier transform (NFT) pump probe spectroscopy is an experimental
  technique
  which allows investigation of the electronic excitation, ionization,
  and unimolecular
  dissociation processes. The NFT
  spectroscopy utilizes ultrafast multiphoton ionization in
  the extreme ultraviolet  spectral range and
  detects the dissociation products of the unstable
  ionized species. In this
  paper, a quantum mechanical description of NFT spectra is suggested,
  which is based on the second order perturbation theory in molecule-light
  interaction and the
  high level ab initio  calculations of CO$_2$ and CO$_2^+$ in the
  Franck-Condon zone.
  The calculations capture the characteristic features of the 
  available experimental NFT spectra of CO$_2$. Approximate
  analytic expressions are derived and used to assign the calculated spectra 
  in terms of participating electronic states and harmonic photon frequencies.
  The developed approach provides a convenient framework within which 
  the origin and the significance of near harmonic and non-harmonic NFT 
  spectral lines 
  can be analyzed. The framework is scalable and
  the spectra of di- and
  triatomic species as well as the dependences on the control parameters
  can by predicted semi-quantitatively. 
\end{abstract}

\maketitle

\section{Introduction}\label{intro}

Chemical transformations, induced by the ultraviolet (UV),
vacuum UV (VUV), and extreme UV (XUV) light in 
carbon dioxide CO$_2$ and carbon dioxide cation CO$_2^+$, are
of considerable importance
for atmospheric, planetary, and interstellar chemistry.
Spectral signatures of the cation CO$_2^+$ were
detected in the Martian atmosphere and the comet comae and
tails.\cite{GC10,HMKA80} 
CO$_2$ is the second common trace gas in the Earth
atmosphere. It is one of the main products of the fossil fuel
burning\cite{NOTE-CO2-0} and its photoabsorption
is used in the UV diagnostics of
high-temperature and high-pressure flames.\cite{JSMOBLDH05} Accurate
knowledge of its low-temperature UV absorption properties would improve the
existing photochemical models of the atmospheres of Mars, which is to a
95\% CO$_2$-based,\cite{FBDD00} and of Titan,\cite{VYC08} in which
CO$_2$ is a minor constituent.

Structure, properties, and photodynamics of carbon dioxide and
carbon dioxide cation are thoroughly
studied.\cite{HERZBERG67,L72,RNMLAVKS83,CJL87,ZG90,BCRRSFW91,SFCCRWB92,SL93,BCEKLRTVW96,LCHHESN00,L08,CLJYP10,GC10,G12A,GB12,G13A,G13B,ASSLONJB13,LCYNJ14,LCGBSNJ14,SGCLNJ14,KRSGJLLV11,ASSG17,KWBHC20}
Nevertheless, their photoreactivity in the gas phase and at catalytic
interfaces remains an area of active research mainly
due to its environmental and technological relevance. For example, 
UV light ultimately destroys CO$_2$ with a unit quantum yield.
This reaction provides a one-step route towards CO$_2$ reduction to carbon
monoxide\cite{G16A} and is in scope of studies on
the negative emission technologies.\cite{TJ02,LNWG18,G17}
From a broader perspective, 
carbon dioxide excited with energetic (UV/VUV/XUV)
photons undergoes a series of 
fundamental photochemical processes typical of 
highly energy loaded molecular systems.
These include non-adiabatic interactions
between Rydberg and valence electronic states,\cite{ASSG17}
roaming dissociation pathways
leading to unusual photochemical products,\cite{B14,LCYNJ14} 
or dissociative VUV photoionization.\cite{YYCZYW19} Understanding the atomistic
and electronic mechanisms of these processes enhances our ability
to address the current major
technological and climate challenges at the molecular level.

\begin{figure}[t]
  \includegraphics[angle=0,scale=0.50]{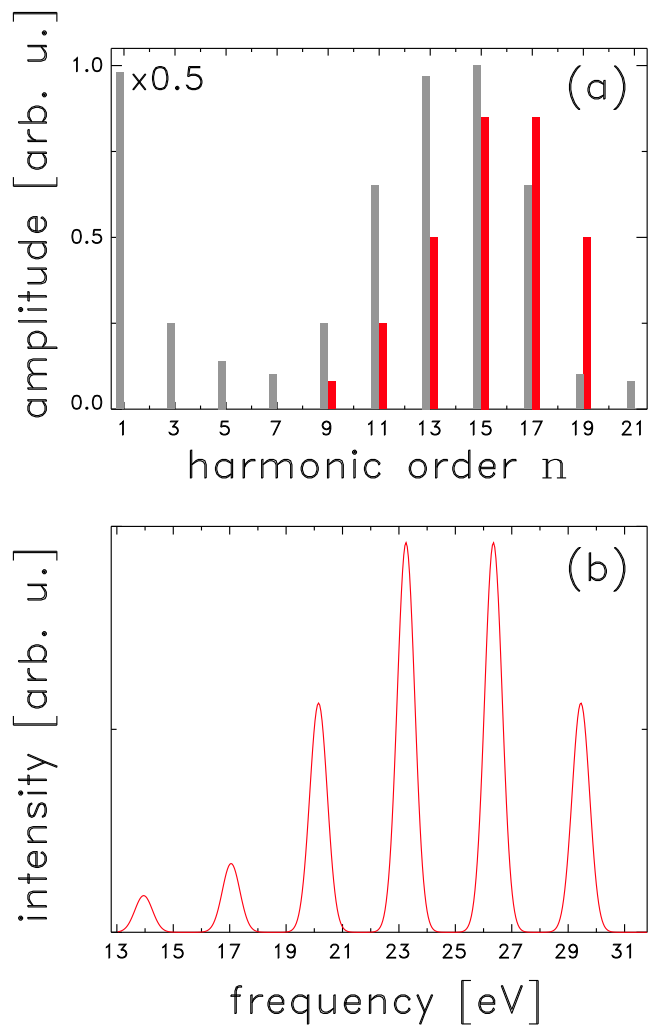}
  \vspace{1.5cm}
  \caption{
    (a) Amplitudes $a_n$ of the harmonics making up the APTs used in the
    calculations [cf. Eq.\ (\ref{eapt2})].
    Red rectangles: Amplitude Set 1
    with only harmonics between $n = 9 $ and $n = 19$ included. 
    Grey rectangles: Amplitude Set 2 modeling the experiment of
    Ref. \onlinecite{OFSNYM14}, but with the harmonic amplitude for $n=1$
    strongly amplified to the value of 2.
        (b) The frequency spectrum of one of the APTs in the calculations. The
     fundamental frequency of
     $\omega_0 = 1.55$\,eV and the amplitude Set 1 [red rectangles in panel
       (a)] are used. The temporal width of the APT envelope is 5\,fs, giving
     the spectral width of the harmonic lines of about 0.80\,eV. 
  }
  \label{apt}
\end{figure}

The rapidly expanding field of attochemistry, utilizing the generation of
ultrashort attosecond-scale XUV pulses, interrogates
the photochemical
processes on the time scale typical for the motion of valence
electrons. Various spectroscopic and time-resolved pump-probe techniques
have emerged, including attosecond spectroscopy\cite{BSJ19,RLN16,SRLG19,CTC17}
and streaking,\cite{PFNB12,CMPY07} 
2D correlation spectroscopy,\cite{LDH20,KWBHC20}
and Raman scattering spectroscopy.\cite{TM02}
The choice of the particular method is 
primarily dictated by the nature of the attosecond dynamics one intends to 
study. Attosecond nonlinear Fourier Transform (NFT) spectroscopy in XUV
is an 
experimental attosecond method which can be directly applied to the
investigation of super-excited ionizing and dissociating molecular
systems.\cite{SKKW04,OYSMNM08,OFSNYM14} With this technique, one 
detects ions resulting from the coherent interaction
between two attosecond pulse trains (APTs) and a molecule excited and ionized
via multi-photon transitions. The spectral content of the APTs 
produced through high-harmonic generation\cite{LBILC94} is
commonly limited to odd harmonics 
of the fundamental driving frequency $\omega_0$ (usually in the infrared).
Figure \ref{apt}(a) provides examples of the intensity distributions of the
harmonics in the APTs used in this work. 
The two APTs propagate along two
interferometric arms with variable delay $\tau$ and are focused on the
interaction region containing only the molecular species of interest.
The NFT spectroscopy is an attosecond
pump-probe detection scheme, with identical
pump and probe APT pulses. The time delay increments in
this technique, determined
by the reproducibility with which one can move the interferometric arms
and control the relative delay, is in the latest experiments
of the order or better than 10\,as.\cite{NSOFHYM09} 
The frequency resolution in the NFT spectroscopy is limited in principle
by the maximum delay $\tau_{\rm max}$
of the interferometer, typically of a few 
femtoseconds,\cite{NSFT09,OFSNYM14} which corresponds to $\sim 1$\,eV. The
excitation energy resolution, determined by the width of the
temporal envelope of the APTs,
is also of the order of 1\,eV [cf. Fig.\ \ref{apt}(b)].

Attosecond
NFT spectroscopy has the potential to provide new information on the
dynamics of photodissociation
of CO$_2$ and CO$_2^+$.\cite{OYSMNM08,OFSNYM14} 
The nonlinear response of CO$_2$ to APT fields
is encoded in the
interferometric autocorrelator employing velocity map imaging (VMI), 
which detects the signals corresponding to the fragment ions, such as  
C$^+$, O$^+$, CO$^+$,  as functions of the time delay $\tau$ between the
two APTs. For each fragment ion type, the signal of an electron/ion
time-of-flight spectrometer such as VMI is
rendered\cite{EP97} as a
two-dimensional (2D) autocorrelation map giving the fragment kinetic energy
distribution for
each time delay $\tau$.
The 2D NFT spectrum, showing the intensity as a function
of the kinetic energy and the NFT frequency $\omega_{\rm NFT}$,
is obtained by a Fourier transform from the $\tau$ domain to the
$\omega_{\rm NFT}$ domain.
Integration over the kinetic energy of the ionic fragment gives a
one-dimensional (1D) NFT spectrum representing the
total ionic yield as a function of $\omega_{\rm NFT}$ and proportional
to the total
population of the electronic states of the ionized molecule dissociating
into the channels containing detected fragments.

The goal of this work is to explore the quantum mechanical aspects of
the experimental NFT spectroscopic technique, 
to construct a theoretical ab initio model for calculation
of the NFT spectra of carbon dioxide, and to identify the NFT spectral
features which can carry information on the chemical rearrangements within 
the molecule or the cation. This study
provides transparent theoretical means
to predict ab initio NFT spectra of small polyatomic molecules
with a modest computational effort. 
We hope that these results can be
used to support and to inspire new experimental campaigns.

In what follows, we concentrate on the
simplest type of the pump-probe signal ---
the total ionic yield after two APTs and
the associated 1D NFT spectra. The extension to 2D spectra is considered
in a separate publication, although we briefly summarize our approach in
appendix \ref{appa}. 
The quantum mechanical description of NFT spectra can be conveniently based
on the general framework, developed by Seel and Domcke\cite{SD91,SD91A} for 
two-pulse time resolved ultrafast ionization spectroscopy
of polyatomic molecules. The laser radiation field is treated classically. 
The interaction between molecule or cation and light
is accounted for using the time dependent perturbation theory.
The ladder of
electronic states of CO$_2^+$ is computed using high level
electronic structure theory including electron correlation.

The analysis of the theoretical NFT spectra of carbon dioxide enables one
to evaluate NFT spectroscopy as a research tool and to discuss the following
questions: (1) How to assign spectral peaks in 1D NFT spectra? (2) Under
which circumstances spectral peaks at non-harmonic NFT frequencies can
develop and what is their significance? (3) How sensitive are NFT spectra
to variations in the fundamental laser frequency $\omega_0$?

The paper is organized as follows: The quantum mechanical approach to NFT
spectra is outlined in Sect.\ \ref{theo1}. This section describes the
physical and chemical aspects
of the APT-induced photoionization of CO$_2$,
summarizes the 
main simplifying assumptions, introduces the Hamiltonian for the system
consisting of the molecule and cation interacting with the laser light,
and relates the 1D NFT spectrum to the solution of the time dependent
Schr\"odinger equation. The ab initio quantum chemical calculations of the
molecular and ionic electronic states are discussed in Sect.\ \ref{abinitio}.
The NFT spectra of CO$_2$ are presented in Sect.\ \ref{spec} and compared with
the available experimental data. The calculated spectra are
assigned in terms of the participating electronic states and the
harmonic orders involved in electronic transitions in Sect.\ \ref{assign}. 
Section \ref{summary} concludes and provides an outlook on the applications
of the developed theory, illustrating how the
dependence of the NFT spectra on the control parameters can be visualized.
Two appendices provide additional information.
In Appendix\ \ref{appa}, the main equations
extending the developed theory to 2D spectra are derived. In
Appendix\ \ref{appb}, 
approximate analytical expressions which support and guide the assignment
of the NFT spectra are presented.

\section{Quantum mechanical approach to NFT signals}
\label{theo1}

\subsection{The photochemical model}
\label{photochem}

\begin{figure}[h!]
  \includegraphics[angle=-90,scale=0.5]{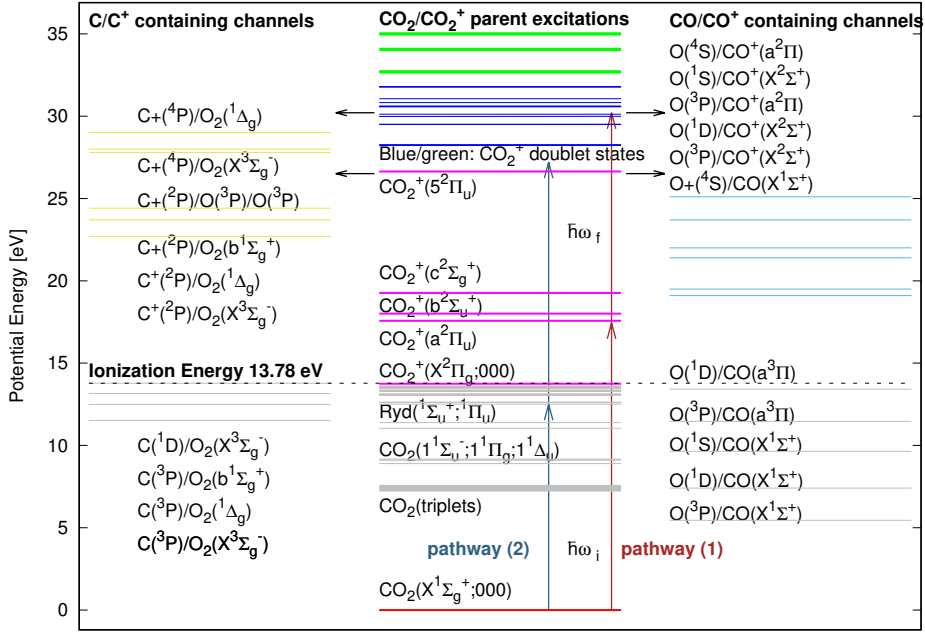}
  \vspace{1.0cm}
  \caption{An overview of
    ab initio electronic spectra of CO$_2$ (mainly singlet states)
    and CO$_2^+$ (doublet states).
    The ground electronic state of CO$_2$ and the electronic 
    states of CO$_2^+$ are calculated in this work and used in the
    quantum mechanical NFT spectral calculations.
    The excited electronic states of CO$_2$,
    shown with gray color, are taken from
    Refs. \onlinecite{ASSG17} and \onlinecite{G17}. 
    The known two-body dissociation channels
    of CO$_2$ and CO$_2^+$, with dissociation products and 
    their electronic states. Brown arrows illustrate photochemical pathway
    of Eq.\ (\ref{chemeq1}); dark blue arrows --- photochemical pathway
    of Eq.\ (\ref{chemeq2}). Black arrows indicate dissociations in the
    electronic states of CO$_2^+$. 
  }
  \label{fig_energies}
\end{figure}

The APT pulses interact with the parent molecule and trigger photoionization
and photodissociation reactions. One can broadly distinguish two major 
photoreaction pathways. In the first one, the parent molecule
$\left({\rm CO}_2\right)_{\rm X}$
in the ground electronic state $\tilde{X}^1\Sigma_g^+$ absorbs a photon with
frequency $\omega_i$ and becomes ionized
to form cation $\left({\rm CO}_2^+\right)^\star$, which 
is further excited with a photon $\omega_f$ into 
dissociative state(s) $\left({\rm CO}_2^+\right)^{\star\star}$: 
\begin{subequations}
\label{chemeq1}
\begin{align}
& \left({\rm CO}_2\right)_{\rm X} + \hbar\omega_i   \rightarrow 
\left({\rm CO}_2^+\right)^\star + e^- \,; \label{chemeq1a}\\
& \left({\rm CO}_2^+\right)^\star + \hbar\omega_f  \rightarrow 
\left({\rm CO}_2^+\right)^{\star\star} \label{chemeq1b}
\end{align}
\end{subequations}
In the second pathway, the parent molecule is 
promoted into an electronically
excited neutral state before ionization: 
\begin{subequations}
\label{chemeq2}
\begin{align}
& \left({\rm CO}_2\right)_{\rm X} + \hbar\omega_i   \rightarrow 
\left({\rm CO}_2\right)^\star\,; \label{chemeq2a}\\
& \left({\rm CO}_2\right)^\star + \hbar\omega_f  \rightarrow 
\left({\rm CO}_2^+\right)^{\star\star} + e^-\,; \label{chemeq2b}
\end{align}
\end{subequations}
This pathway is akin to the one explored recently by Adachi et al. 
in the experimental study of the 
ultrafast ionization spectroscopy of CO$_2$.\cite{ASSG17}
An overview of the neutral and ionic
electronic states mediating different photoreaction pathways
is given in Fig.\ \ref{fig_energies}.  
The photoexcitations via Eq.\ (\ref{chemeq1}) are shown with brown arrows
and via Eq.\ (\ref{chemeq2}) --- with dark blue arrows. 
In either pathway, the unstable cation 
$\left({\rm CO}_2^+\right)^{\star\star}$ dissociates
into the arrangement channels
containing fragment ions, for example O$^+$, CO$^+$, or C$^+$:
\begin{equation}
\label{chemeq3}
\left({\rm CO}_2^+\right)^{\star\star}  \rightarrow 
\left\{\begin{array}{ll}
    {\rm O}^+ + {\rm CO}+ e^- \, ; \\
{\rm CO}^+ + {\rm O}+ e^- \, ; \\ 
{\rm C}^+ + {\rm OO}+ e^- \, . 
\end{array}\right.
\end{equation}
The fragment ions are ultimately
detected using VMI. The dissociation threshold relevant for the production
of O$^+$/CO$^+$ fragment ions is close to 19.0\,eV 
(see Ref.\ \onlinecite{HBV80} and Fig.\ \ref{fig_energies} in which thresholds
for various O$^+$/CO and O/CO$^+$ channels are marked on the energy scale
above the ground state of the neutral CO$_2$). The
appearance of C$^+$ ions is established to occur between 25.0\,eV and
30.0\,eV and is attributed to the 3-body dissociation;\cite{HBV80,M94}
in Fig.\ \ref{fig_energies}, the C$^+$/O/O channel is located
close to 25\,eV. As an aside, we note that recent experiments of
Lu et al. demonstrated that CO$_2$ can decompose into the 2-body channel C and
O$_2$.\cite{LCYNJ14} This might suggest that a similar
2-body arrangement channel C$^+$/O$_2$, lying upwards of 17.0\,eV, 
could be detected, too. However, we found no
published experimental result so far.

The two-photon
excitations in these reactions can be either concerted or
sequential. General expressions for the NFT signals, 
discussed in Sects.\ \ref{theo1_sig} and \ref{theo1_pt},
account for both pathways. The ab initio
model in Sect.\ \ref{abinitio} is constructed for the reactions of Eqs.\
(\ref{chemeq1}) and (\ref{chemeq3})
which constitute the relevant pathway under the experimental
conditions of Ref.\ \onlinecite{OFSNYM14}. 

The above reaction schemes illustrate the main assumptions made
in this work:\newline
\noindent
{\bf A1. Excitations in the Franck-Condon zone only}.
Equations (\ref{chemeq1})  and (\ref{chemeq2}) imply that all interactions
with photons take place in the Franck-Condon zone, before
either the neutral CO$_2$ or the cation CO$_2^+$ start to decompose. This
assumption is justified: Both the APT durations and the pump probe delay
times in the NFT experiments are smaller than
10\,fs\cite{OYSMNM08,OFSNYM14} and
are therefore substantially shorter than
the characteristic times of vibrational motion in CO$_2$ or CO$_2^+$.
Judging by the resonance lifetimes of $\sim 50 - 100$\,fs
calculated for CO$_2$,\cite{G12A,G13B} 
the dissociation reactions are expected to unfold
on a much longer time scale. This has two implications: First, the ionization
of the neutral dissociation fragments, such as CO, O, or C,
makes no contribution to the observed
NFT signal and can be ignored. Second, the electronic states 
contributing to the total ion yield (or the 1D
NFT spectrum), as well as the transition matrix elements between them,
can be found from the quantum chemical
calculations limited to
the Franck-Condon zone. This simplification is used in Sect.\ \ref{abinitio}
to set up the ab initio model. 

\noindent
{\bf A2. Single ionizations only}.
The reaction schemes in Eqs.\ (\ref{chemeq1}) and (\ref{chemeq2})
involves only single ionizations.
Double ionization of carbon dioxide is also possible and has been
extensively investigated.\cite{SKH77,M94,ACFLPRSV10,ELBXK16}
The cross section grows with energy
in excess of threshold located 37.3\,eV above minimum of CO$_2$.
However, the ratio of
CO$_2^{2+}$ to CO$_2^+$ does not exceed 2\% even 30\,eV above
threshold,\cite{SKH77,M94} and can be neglected in experiments operating
APTs with sum frequencies below 70\,eV or so. 
This assumption is also consistent with the
mass spectroscopic measurements of Ref.\ \onlinecite{OFSNYM14} which indicate
very low intensities for peaks corresponding to the doubly ionized species.

\noindent
{\bf A3. Two-photon processes only}.
The reactions in Eqs.\
(\ref{chemeq1}) and (\ref{chemeq2}) consume only two photons. This is
in line with the second order perturbation theory in the molecule-light
interaction which we use to 
find the time dependent
excitation and ionization amplitudes. The perturbation theory
is known to be reliable for ultrashort laser pulses.\cite{SSD89} The
role of higher order processes in the NFT spectroscopy of CO$_2$
was discussed,\cite{OFSNYM14} but no conclusive evidence was found so far. 
An extension to nonperturbative treatment of excitations and ionizations
can be made (see, for example, Ref.\ \onlinecite{SD91}), but is outside
the scope of this work.

With these assumptions, we seek to develop a minimum theoretical description
      adequate for a quantitative analysis of experimental 1D
      NFT spectra of photoionizing small polyatomic molecules --- taking
      CO$_2$ as an example.

\subsection{The Hamiltonian}  
\label{theo1_ham}
Quantum mechanical theory of NFT spectra in the setup involving
ultrafast time resolved ionization of polyatomic molecules is based on the
approach developed in the seminal papers by Seel and Domcke.\cite{SD91,SD91A} 
The electronic basis
includes the ground electronic state
$\left|\phi_0\right.\rangle$ of CO$_2$; a set of
excited electronic states of neutral
CO$_2$, $\{\left|\phi_\alpha\right.\rangle\}$; 
the one-electron continuum states $\left|\psi^e_k\right.\rangle$,
corresponding to the photoelectron
kinetic energy $E_k$; and a set of ion core states
$\{\left|\phi_j^+\right.\rangle\}$ of CO$_2^+$.
Direct products $\left|\phi_j^+\psi^e_k\right.\rangle$
define the ionization continua in the model.\cite{NOTE-NFTS-01}
The choice of the electronic basis is governed by the assumption {\bf A2} of
Sect.\ \ref{photochem} --- only single ionizations are considered. 

The molecular Hamiltonian in this basis has the following
form:
\begin{equation}
  \label{hamm}
  H_M = \left|\phi_0\right.\rangle H_0 \langle \left. \phi_0 \right| +
  \sum_\alpha\left|\phi_\alpha\right.\rangle
    H_\alpha\langle\left.\phi_\alpha\right| + 
  \sum_j\int_0^\infty d\,E_k\, \left|\phi^+_j\psi^e_k\right.\rangle
  \left(H_j + E_k\right)\langle\left.\phi^+_j\psi^e_k\right| \, .
\end{equation}
The electronic basis states, both neutral and ionic,
are treated as diabatic. Possible non-adiabatic off-diagonal interactions
between them are suppressed and will be explicitly considered in a
separate publication on 2D NFT spectra. 

The Hamiltonians $H_0$, $H_\alpha$, and $H_j$ describe the vibrational
dynamics in 
the electronic ground state of CO$_2$, in the excited electronic states
of CO$_2$, 
and in the electronic states of CO$_2^+$, respectively. Vibrational
eigenstates in each electronic state are given by:
\begin{eqnarray*}
\label{hamvib}
H_0 |0\rangle & = & \epsilon_0 |0\rangle \, ;\\
H_\alpha |v_\alpha\rangle & = & \epsilon_{v\alpha} |v_\alpha\rangle \, ;\\
H_j |v_j\rangle & = & \epsilon_{vj} |v_j\rangle \, ;
\end{eqnarray*}
The vibrational energies $\epsilon_{v\alpha}$ and $\epsilon_{vj}$
are measured with respect to the energy $\epsilon_{0}$ of the
ground vibrational state in the state $\tilde{X}^1\Sigma^+_g$
of CO$_2$. For example, the
energy $\epsilon_{vj = 1}$ of the ground vibrational state in the ground 
electronic
state $\tilde{X}^2\Pi_g$ of CO$_2^+$ is approximately equal to 13.8\,eV, the
ionization energy of CO$_2$ [cf. Fig.\ \ref{fig_energies}].

The external electric field ${\cal E}(t)$ 
is comprised of two APT fields, one of which is
delayed by a time $\tau$: 
\begin{equation}
\label{etot}
{\cal E}(t) = {\cal E}_{\rm APT}(t,0) + {\cal E}_{\rm APT}(t,\tau) 
\, .
\end{equation}
The time profile of each APT field is determined by the envelope
function $L_n(t,\tau)$:\cite{OFSNYM14}  
\begin{equation}
\label{eapt1} 
{\cal E}_{\rm APT}(t,\tau)  = 
\sum_{n = n_1}^{n_2}{}^{'} a_n L_n(t,\tau) e^{-i\omega_n (t-\tau)}\, .
\end{equation}
Here $n$ is the harmonic order (the primed
sum runs only over odd orders); $\omega_n$
are the harmonic frequencies; $a_n$ is the amplitude of the $n$-th harmonic
in the APT. Gaussian time envelope is often considered,
\begin{equation}
\label{eapt2}
L_n(t,\tau) =  \left(\frac{P}{\pi}\right)^{1/2} \frac{1}{T}\,
e^{-P(t-\tau)^2/T^2}\, , 
\end{equation}
and will be used in the numerical calculations in this work. In this
expression, $T$ is the FWHM of the APT Gaussian envelope and  
$P=4\ln 2$. The pump-probe time delay $\tau$ is the 
time interval between the centers of the Gaussian envelopes
of the two APTs. The APT spectral shape for the pulse of an experimentally
realistic\cite{OFSNYM14} duration of $T = 5$\,fs is shown in Fig.\
\ref{apt}(b). The widths of the individual harmonic peaks are slightly below
1\,eV and therefore are comparable to the fundamental frequency $\omega_0$
(which is 1.55\,eV in the figure).

The interaction between the molecule and the external electric field
${\cal E}(t)$ includes the following components capable of 
describing
the pump-probe dynamics along the photoreaction pathways of
Eqs.\ (\ref{chemeq1}) and (\ref{chemeq2}):\cite{NOTE-NFTS-03} 
\begin{enumerate}
\item {\bf Photoreaction pathway of Eq.\ (\ref{chemeq1}).} 
The interaction term describing this pathway comprises two components: 
\begin{equation}
  \label{w1}
  W_{1}(t) = W_{1A}(t) + W_{1B}(t) \, .
\end{equation}
with one of them mediating ionization out of the ground state of CO$_2$,
\begin{equation}
\label{w1a}
W_{1A}(t) = - \sum_i \int_0^\infty d\,E_k\,\left|\phi^+_i\psi^e_k\right.\rangle
\mu_{i0}(E_k){\cal E}(t)\langle\left.\phi_0\right|\, +\, h.c. \, ,
\end{equation}
and the other allowing optical excitations between states of the
free ionic core: 
\begin{equation}
\label{w1b}
W_{1B}(t) = - \sum_{i,j\ne i}\left|\phi^+_j\right.\rangle \mu_{ji}{\cal E}(t)
\langle\left.\phi^+_{i}\right|
\int_0^\infty d\,E_k\,\left|\psi^e_k\right.\rangle
\langle\left.\psi^e_k\right|\, +\, h.c. 
\end{equation}
In these expressions, 
$\mu_{i0}(E_k)$ is the ionization dipole moment, and
$\mu_{ji}$ is a transition dipole moment (TDM)
between ionic states $j$ and $i$. 

\item {\bf Photoreaction pathway of Eq.\ (\ref{chemeq2}).} The interaction term
is structurally similar to the above:
\begin{equation}
  \label{w2}
  W_{2}(t) = W_{2A}(t) + W_{2B}(t) \, ,
\end{equation}
with one term describing optical excitations in the neutral molecule,
\begin{equation}
\label{w2a}
W_{2A}(t) = - \sum_{\alpha}\left|\phi_\alpha\right.\rangle
\mu_{\alpha 0}{\cal E}(t)
\langle\left.\phi_{0}\right|\, +\, h.c. \, ,
\end{equation}
and the other giving rise to the 
ionization from the state $\left|\phi_\alpha\right.\rangle$ of CO$_2$:
\begin{equation}
\label{w2b}
W_{2B}(t) = - \sum_{j\alpha}
\int_0^\infty d\,E_k\,\left|\phi^+_j\psi^e_k\right.\rangle
\mu_{j\alpha}(E_k){\cal E}(t)\langle\left.\phi_\alpha\right|\, +\, h.c. \, .
\end{equation}
Here $\mu_{\alpha 0}$ are the TDMs for optical excitations in the neutral
CO$_2$ from the
ground electronic state $\tilde{X}^1\Sigma^+_g$, and 
the ionization dipole moments
$\mu_{j\alpha}(E_k)$ are now defined
for a given pair of the ionic state $j$ and the neutral state $\alpha$. 
\end{enumerate}
The total Hamiltonian governing the dynamics in the
time dependent electric field ${\cal E}(t)$ is given by the sum of the
molecular Hamiltonian [Eq.\ (\ref{hamm})] and the interaction with laser
field: 
\begin{equation}
  \label{htot}
  H_{TOT}(t) = H_M + W_{1}(t) + W_{2}(t)\, .
\end{equation}

The TDMs between diabatic states are treated in the Condon approximation
consistent with the diabatic representation. For the ionization step, the
dipole moments $\mu_{i0}(E_k)$ and $\mu_{j\alpha}(E_k)$ 
depend on the photoelectron
kinetic energy $E_k$ and are subject to the boundary
condition $\mu(E_k) \rightarrow 0$ as $E_k \rightarrow \infty$. 
Following
Seel and Domcke,\cite{SD91} we approximate
this dependence by a simple step function, e.g.
\begin{equation}
  \label{muek}
  \mu_{i0}(E_k) =
  \begin{cases} \mu_{i0}, & \mbox{if }  0 \le E_k \le E_k^{\rm max} \\
    0, & \mbox{if }   E_k > E_k^{\rm max} \end{cases}
\end{equation}
The cutoff energy $ E_k^{\rm max}$ is a parameter of the calculation
controlling the width of the photoelectron spectrum in a given ionic state.

\subsection{NFT signals}  
\label{theo1_sig}
The NFT signal calculated in this work is the
total yield $I_{\rm ion}(\tau)$ of the detected fragment ion
[for example, C$^+$, O$^+$, or CO$^+$, cf. Eq.\ (\ref{chemeq3})].
This 1D signal is proportional to the total
population of the ionic electronic states dissociating to produce the
detected ion. The ladder model in Fig.\ \ref{fig_energies},
approximating the electronic spectrum of CO$_2^+$ with
ab initio energies at the
Franck-Condon point, can be taken as a starting point for a calculation of
$I_{\rm ion}(\tau)$. 
After the pump and probe pulses, the system, consisting of 
CO$_2^+$ and a photoelectron, is left in the state
$\Psi_I(t|E_k,\tau)$. The population $p_f$ of a dissociative final state 
$|f\rangle$ of CO$_2^+$ is given by 
\begin{equation}
\label{pop0}
p_f(E_k,\tau) = \left|{\langle \phi_f^+\psi^e_k
  |\Psi_I(t\rightarrow\infty|E_k,\tau)\rangle}\right|^2\, .
\end{equation}
The total ion yield $I_{\rm ion}(\tau)$ is proportional to this population
integrated over the photoelectron kinetic energy $E_k$ and 
summed over all final states having energies $\epsilon_f$
above the appearance threshold $A_{\rm ion}$ of the detected ion:
\begin{equation}
  \label{ion_t}
  I_{\rm ion}(\tau) = \sum_{f:\, \epsilon_f > A_{\rm ion}}
    \int_0^\infty d\,E_k\, p_f(E_k,\tau) \, .
\end{equation}
Fourier transform of  $I_{\rm ion}(\tau)$
gives the
frequency domain 1D NFT spectrum:
\begin{equation}
  \label{ion_w}
  I_{\rm ion}(\omega_{\rm NFT}) =
  \int_{-\tau_{\rm max}}^{\tau_{\rm max}} d\tau I_{\rm ion}(\tau)
  e^{i\omega_{\rm NFT}\tau}
\end{equation}
The maximum time delay $\tau_{\rm max}$
defines the spectral resolution in the NFT frequency $\omega_{\rm NFT}$.

NFT signals of CO$_2$ are also reported as 2D maps 
$I_{\rm ion}(\epsilon_{\rm kin},\tau)$, in which the kinetic
energy distribution of the recoiling fragment ions
is measured for different time delays. The NFT observable in this case
is directly related to the 
dissociation dynamics in the ionic states, and the description of the signal
requires basic elements of scattering theory. The scattering
approach to the 2D NFT signals is outlined in 
Appendix \ref{appa}. 
Numerical applications are considered in a separate publication. 

\subsection{Total ion yield via time dependent perturbation theory}  
\label{theo1_pt}

Evaluation of the NFT signals described with Eqs.\
(\ref{pop0}) and (\ref{ion_t}), as well as with Eqs.\
(\ref{i_et}) and (\ref{i_t2}), requires the time dependent molecular
wave function $\left|\Psi_I(t;E_k,\tau)\right.\rangle$
in the interaction representation. We expand it in the electronic basis
introduced in Sect.\ \ref{theo1_ham}: 
\begin{equation}
\label{psit}
\left|\Psi_I(t;E_k,\tau)\right.\rangle = \chi_0(t)\left|\phi_0\right.\rangle + 
\sum_\alpha \chi_\alpha(t) \left|\phi_\alpha\right.\rangle + 
\sum_j\int_0^\infty d\,E_k \chi_j(t;E_k,\tau)\,
\left|\phi^+_j\psi^e_k\right.\rangle\, ,
\end{equation}
Here the coefficients $\chi_0(t)$ and $\chi_\alpha(t)$ are
the nuclear wave functions in the 
ground (index 0) and excited (indices $\alpha$)
electronic states of the neutral molecule; $\chi_j(t;E_k,\tau)$
are the nuclear wave functions of the molecular ion depending
on the photoelectron kinetic energy $E_k$ and, through the APT fields,
on the pump probe time delay $\tau$. 
The initial state of the molecule is assumed to be the vibrational ground
state in the electronic ground state,
$\left|\phi_0\right.\rangle\left|0\right.\rangle$.

The coefficient $\chi^{(2)}_j(t;E_k,\tau)$ for the CO$_2^+$ ion in one of
the final dissociative states is evaluated using the
second order time dependent
perturbation theory in the interactions $W_1(t)$ and $W_2(t)$.
This is in line with the assumption {\bf A3} of
Sect.\ \ref{photochem}, so that the theoretical description focuses on the 
two-photon processes. Perturbation theory allows us to concentrate on the
observable ion signal and bypass a rigorous description of the ejected electron
which has not been detected
in the NFT spectroscopic experiments on CO$_2$. In fact, 
the photoelectron dynamics is 
very rich in carbon dioxide,\cite{DO79}
and was a subject of detailed theoretical and
experimental studies on the attosecond time scale.\cite{KRSGJLLV11,KWBHC20}

The
contributions of the interactions with laser fields
to the wave functions (and to the pump probe
amplitudes) are additive. Considering the interaction $W_1(t)$, the
second order nuclear wave function is given by: 
\begin{equation}
  \label{chij}
  \chi^{(2)}_j(t;E_k,\tau) = \frac{1}{i^2}\sum_{i\ne j}
  \int_{-\infty}^t dt_1\int_{-\infty}^{t_1} dt_0\, e^{iH_{j}t_1}
  \left[\mu_{ji}{\cal E}(t_1)\right] e^{iE_kt_0}e^{-iH_{i}(t_1-t_0)}
  \left[\mu_{i0}(E_k){\cal E}(t_0)\right]e^{iH_0t_0}\left|0\right.\rangle \, .
\end{equation}
We use $\hbar = 1$ throughout the paper. 
The amplitude $\chi^{(2)}_j$ 
describes ionization of the molecule in the vibrationless
ground state $|0\rangle$ into an ionic state $i$ at time $t_0$,
and a subsequent optical excitation of the state $i$ into the final
state $j$ at time $t_1$ [photoreaction via the Eq.\ (\ref{chemeq1})].
The pump probe experiment is represented as a linear
combination of photoionization/excitation events separated by the temporal
delay of length $\tau$. Projecting this function onto the final
vibrational state $\langle v_j|$ and taking the limit
$ t\rightarrow \infty$ gives the asymptotic second order pump probe amplitude
due to the interaction $W_1(t)$:  
\begin{eqnarray}
\label{cvj}
c_j(v_j;E_k,\tau) & = &  \frac{1}{i^2}\sum_{i\ne j}\sum_{vi}\mu_{ji}
\langle v_j | v_{i}\rangle \mu_{i0}(E_k) \langle v_{i}|0\rangle \nonumber \\
&\times& \int_{-\infty}^\infty dt_1 e^{i(\epsilon_{vj} - \epsilon_0 + E_k) t_1}
{\cal E}(t_1)
\int_{-\infty}^{t_1} dt_0\, e^{-i(\epsilon_{vi} - \epsilon_0 + E_k)(t_1-t_0)}
{\cal E}(t_0)\, .  
\end{eqnarray}
The pump probe amplitudes $c_j(v_j;E_k,\tau)$ depend on the time delay $\tau$
via the time dependence of the laser field ${\cal E}(t)$, cf. 
Eqs.\ (\ref{etot}) and Eqs.\ (\ref{eapt1}) and (\ref{eapt2}). 
The coefficients $\langle v_{i}|0\rangle$ in Eq.\ (\ref{cvj})
are the expansion coefficients of the initial
vibrational state $|0\rangle$ in the vibrational states of the intermediate
ionic states, $|v_{i}\rangle$; the coefficients 
$\langle v_j | v_{i}\rangle$ are the projections of the 
intermediate ionic vibrational states onto the final state $|v_{j}\rangle$.

Repeating the same steps for the
interaction $W_2(t)$, one finds another contribution to the pump
probe amplitude due to an optical
excitation of the molecule in the vibrationless
ground state $|0\rangle$ into a neutral state $\alpha$ at time $t_0$,
and a subsequent ionization of the state $\alpha$ into the final
state $j$ at time $t_1$ [photoreaction via the Eq.\ (\ref{chemeq2})]: 
\begin{eqnarray}
\label{cvjb}
d_j(E_k,v_j,\tau) & = &  \frac{1}{i^2}\sum_{\alpha > 0}\sum_{v\alpha}
\mu_{j\alpha}(E_k)\langle v_j | v_{\alpha}\rangle 
\mu_{\alpha 0}\langle v_{\alpha}|0\rangle \nonumber \\
&\times& \int_{-\infty}^\infty dt_1 e^{i(\epsilon_{vj} - \epsilon_0 + E_k) t_1}
{\cal E}(t_1)
\int_{-\infty}^{t_1} dt_0\, e^{-i(\epsilon_{v\alpha} - \epsilon_0)(t_1-t_0)}
{\cal E}(t_0)\, .  
\end{eqnarray}
The coefficients $\langle v_{\alpha}|0\rangle$ and 
$\langle v_j | v_{\alpha}\rangle$
in this equation are similar to the respective coefficients in 
Eq.\ (\ref{cvj}), but refer to the projections of the vibrational states in
the excited electronic states of the neutral molecule. 
The total pump probe amplitude $b_j(E_k,v_j,\tau)$
is given by the sum of the amplitudes for the two pathways: 
\begin{equation}
\label{fvj}
b_j(E_k,v_j,\tau) = c_j(E_k,v_j,\tau) + d_j(E_k,v_j,\tau) 
\end{equation}
The ionic signal $I_{\rm ion}(\tau)$, defined in the Eq.\ (\ref{ion_t}), is
proportional to the population $p_f$ in the final dissociative states
$|f\rangle$ capable of producing the detected ion, 
i.e. to the square of the pump probe amplitudes $|b_f|^2$
collected over vibronic levels $\epsilon_{vf}$ lying above the appearance
threshold $A_{\rm ion}$: 
\begin{equation}
\label{ion_t2}
  I_{\rm ion}(\tau) = \sum_{f}\sum_{\epsilon_{vf} > A_{\rm ion}}
    \int_0^\infty d\,E_k\,\left|b_f(E_k,v_f,\tau)\right|^2  \, .
\end{equation}
This is a working expression in the calculations discussed in Sect.\
\ref{results}. It is valid even if different electronic states interact
non-adibatically\cite{CIBOOK04} --- the vibronic energies $\epsilon_{vf}$ in
this case should refer to adiabatic rovibronic eigenstates. 

\section{Results and discussion}
\label{results}

\subsection{Ab initio calculations of CO$_2$ and CO$_2^+$}
\label{abinitio}

The challenge in modeling NFT spectra for
polyatomics is to construct an adequate (preferably ab initio)
model of electronic states of the neutral molecule and the molecular ion.
Strictly speaking, potential energy surfaces depending on all three internal
coordinates are needed to find the eigenstates $\epsilon_{vj}$ and to
calculate even the total ion yield.  
Due to a multiphoton nature of the NFT technique and the XUV
harmonics used for excitation and ionization, this can easily become a
formidable task because tens or even hundreds (interacting)
electronic states might be needed.\cite{ASSG17}

In this work, we use the assumption {\bf A1} of
Sect.\ \ref{photochem} and calculate the electronic energy levels,
as well as the dipole moments for the ionization and optical excitation steps,
at the equilibrium
geometry of the ground electronic state of the neutral molecule.
The symmetry group of this Franck-Condon point is
$D_{\infty h}$. In the electronic structure
calculations, it is rendered as $D_{2h}$. 

The ab initio calculations are further simplified by concentrating the analysis
on the photoreaction pathway of Eq.\ (\ref{chemeq1}), so that ionizations
take place directly from the ground electronic state of CO$_2$. This
relieves us of having to calculate the densely spaced
mixed Rydberg-valence states of the neutral molecule close to the
ionization threshold.\cite{ASSG17}
The photoreaction pathway of Eq.\ (\ref{chemeq1}) is indeed likely to make
the main contribution to the observed NFT signals of CO$_2$.\cite{OFSNYM14} 
Carbon dioxide is transparent up to about 6.20\,eV (where the absorption
still remains extremely weak),\cite{O71,CLP92,CLP05}
and the first strong absorption band
is observed near 11.08\,eV.\cite{RMSM71,CJL87} The molecule ionizes
at 13.8\,eV, while
the harmonics which substantially contribute to the
APTs considered in the present calculations 
carry excitation energies of more than 14.0\,eV
[cf. Fig.\ \ref{apt}(b)]. Possible spectral signatures of the
complementary pathway of Eq.\ (\ref{chemeq2}) will be indicated in the
discussion of the results. 

The augmented correlation consistent polarized valence quadrupole
zeta (aug-cc-pVQZ) basis set due to Dunning is used for all atoms. Energies
of CO$_2$ and CO$_2^+$  are calculated at the internally-contracted
multireference configuration interaction singles and doubles (MRD-CI) level,
based on state-averaged full-valence complete active space self-consistent
field (CASSCF) calculations with 16 electrons in 12 active orbitals and
6 electrons in three fully optimized closed-shell inner orbitals.
Active space in CASSCF comprises orbitals 
$2\sigma_u - 4\sigma_u$, $3\sigma_g-5\sigma_g$,$1\pi_u-2\pi_u$,
and $1\pi_g$. 
In the MRD-CI step, all 16 valence electrons are correlated.
The Davidson correction is applied in order to account for higher-level
excitations and size extensivity. The singlet ground electronic state of CO$_2$
and a series of doublet states of CO$_2^+$ are calculated with this
setup. 
The MRD-CI calculations for the ion are performed using the molecular orbitals
of the neutral molecule in order to simplify the evaluation of ionization
matrix elements. All ab initio calculations are carried
out with the MOLPRO package.\cite{WKKMS12,MOLPRO-FULL-09}

More than 50 doublet
electronic states of CO$_2^+$ are calculated and assigned. The
full list of converged states includes $1-4\,^2\Sigma_g^+$;
$1-7\,^2\Sigma_u^+$; $1-6\,^2\Sigma_g^-$; $1-6\,^2\Sigma_u^-$; $1-10\,^2\Pi_g$;
$1-10\,^2\Pi_u$; $1-4\,^2\Delta_g$; and $1-4\,^2\Delta_u$.
These states span the energy range from 13.8\,eV to 32.5\,eV above the minimum
of the ground electronic state of the neutral CO$_2$ molecule.

The ionization dipole
moments $\mu_{i0}$ between the states of CO$_2^+$ and the ground electronic
state of CO$_2$ are evaluated for the CI vectors
calculated in the common basis of the molecular orbitals of the
neutral CO$_2$. 
For a given ionic state $i$,
the photoionization dipole moment $\mu_{i0}$ reads as 
\begin{equation}
\label{muj0}
\mu_{i0} = \sum_s {\rm d_{ks}}(E_k) x^s_{i0} \, .
\end{equation}
It is a sum of products of
bound-free one-electron dipole integrals ${\rm d_{ks}}(E_k)$
which depend on the
photoelectron kinetic energy $E_k$, and the 
\lq spectroscopic factors' $x^s_{i0}$ defined as\cite{MLWD06}
\begin{equation}
\label{xsj0}
x^s_{i0} = \langle \phi_i|\hat{c}_s|\tilde{X}^1\Sigma^+_g\rangle\, .
\end{equation}
where  $\hat{c}_s$  stands for the 
annihilation operator for the orbital $s$, and $\phi_i$ is the electronic
wave function of the cation state $i$. The spectroscopic factors are calculated
for the CI vectors of CO$_2$ and CO$_2^+$
using the algorithm proposed by W. Eisfeld.\cite{E05} Note that spin selection
rules are relaxed in photoionization, and higher multiplicity spin states of
CO$_2^+$ (e.g. quartets) can in principle be ionized, too. Test calculations
performed for the four lowest quartet states $1,2^4\!A'$ and $1,2^4\!A''$
demonstrate, however, that the spectroscopic factors for these ionizations
are small.

Optical transitions between the states of the ion require interstate TDMs
$\mu_{ji}$.
Components $(\mu_x,\mu_y,\mu_z)$ of the TDM vector are calculated at the high
symmetry Franck-Condon point.
The coordinate axes in these calculations are chosen such
that $z$  runs along the molecular figure axis, while 
$x$ and $y$ are orthogonal to $z$. The ab initio TDMs are
computed at the CASSCF level of theory. Previous calculations
demonstrate\cite{G13A} that the difference between the TDMs calculated using
CASSCF and MRD-CI methods is less than 10\%.

At the high symmetry Franck-Condon geometry, the
selection rules effectively reduce the
number of accessible
optically bright states; for many pairs of symmetry species, the $x$, $y$,
and $z$ components of the 
TDMs $\mu_{ji}$ vanish. For example, for the five 
states
$1\,^2\Pi_g$, $1\,^2\Pi_u$, $1\,^2\Sigma_u^+$, $1\,^2\Sigma_g^+$,
and  $5\,^2\Pi_u$ (they have the largest 
photoionization probabilities), the
optical transitions are driven by very few 
non-zero TDM components:
\begin{equation}
\label{tdm_matrix}
\left(
\begin{array}{ccccc}
  1^2\,\Pi_g & \mu_z      & \mu_{x,y}       & -               & \mu_z \\
             & 1^2\,\Pi_u & -               & -               & -     \\
             &            & 1^2\,\Sigma_u^+ & \mu_z           & -     \\
             &            &                 & 1^2\,\Sigma_g^+ & -     \\
             &            &                 &                 & 5^2\,\Pi_u 
\end{array}
\right)
\end{equation}
In this symbolic
representation, the TDM components mediating transitions between states
on the main diagonal of the matrix are enumerated as off-diagonal
\lq matrix elements'. The resulting \lq matrix' is sparse. 

Consequently, only a subset of all
calculated states of CO$_2^+$ are included in the quantum mechanical
calculations of the NFT spectra. This subset
is shown in Table \ref{abinitio_en} and in Fig.\ \ref{fig_energies}. 
The accuracy of the ab initio 
calculations can be judged (for the first several states)
by comparison with the
known experimental energies,\cite{WRLS88}
also shown in Table \ref{abinitio_en} where
available. The states of CO$_2^+$ in Table \ref{abinitio_en}
are selected as follows: (1) Five  
states with the largest photoionization probabilities
$|\mu_{i0}|^2$ referenced in Eq.\ (\ref{tdm_matrix}). 
They are
shown above the upper horizontal line in Table \ref{abinitio_en} and as
magenta lines in  Fig.\ \ref{fig_energies}. 
(2) Thirteen states having strong ($> 0.2$\,D) TDMs with the above 5
preselected ionization states are additionally included in the model.
They are shown with blue color in  Fig.\ \ref{fig_energies}.
(3) For several states, assigned $n^2\Sigma_g^+$, 
  $n\,^2\Pi_u$ and $n\,^2\Sigma_u^+$
and shown
below the lower horizontal line in Table \ref{abinitio_en}, convergence at
the MRD-CI level was not achieved. Their energies were shifted in the
quantum mechanical calculations to the values shown in parenthesis. These
states are shown as green lines in Fig.\ \ref{fig_energies}.
These states collectively represent the electronic states
lying outside the energy range covered by the present
ab initio calculations.

\subsection{Calculated and experimental NFT spectra}
\label{spec}

The NFT signals are calculated using Eqs.\ (\ref{ion_w}) and (\ref{ion_t2}), 
and only the pump probe amplitude
$c_j(E_k,v_f,\tau)$ of Eq.\ (\ref{cvj}) is included. 
The vibrational excitations are suppressed for all states,
so that each electronic state contributes one energy $\epsilon_{0j}$
coinciding with the ab initio energy for this state
in Table \ref{abinitio_en}. 

It is assumed that ions in Eqs.\ (\ref{chemeq3})
are produced with the 
appearance thresholds of $A_{\rm C+} = 30.0$\,eV and
$A_{\rm O+} = A_{\rm CO+} = 19.0$\,eV.
All electronic states with
energies above the appearance threshold $A_{\rm ion}$ contribute
to the fragment ion generation. As explained in Sect.\ \ref{photochem},
the appearance thresholds for O$^+$ and CO$^+$ are close to the known
channel dissociation thresholds shown in Fig.\ \ref{fig_energies}, while
$A_{\rm C+}$
is taken substantially higher than the lowest dissociation energy
of 23\,eV, at which the production of C$^+$ ions becomes energetically
allowed.

Two sets of APTs, shown in Fig.\ \ref{apt}(a),  are
used in the calculations. Set 1 (red boxes)
spans the harmonic orders from $n = 9$ to
$n = 19$, with a standard choice of
harmonic amplitudes of the components.\cite{WSG08,PCBGSCCMK09} 
Set 2 covers harmonic orders from $n = 1$
to $n = 21$, as reported for the APTs in Ref. \onlinecite{OFSNYM14}.
Compared to Set 1, it incorporates contributions
from low harmonic orders $n \le 7$. We also artificially amplified the
amplitude for $n=1$. Such an amplitude distribution is valuable because it
describes realistic
experimental conditions under which the fundamental $\omega_0$ or the
adjacent odd harmonics 
are not fully suppressed e.g. by
dichroic mirrors, filters, or harmonic separators. The additional change in
the amplitude for $n=1$ on top of the reported amplitudes, helps to
rationalize the intensity of this harmonic peak in the measured NFT spectrum. 

The ionization transition
matrix elements $\mu_{i0}(E_k)$ between the ground electronic state of
CO$_2$ and the first five ionic states in Table\ \ref{abinitio_en}
are assumed to be constant up to the maximum cutoff energy 
$E_k^{\rm max}$
[cf. Eq.\ (\ref{muek})], and negligible beyond this energy; 
$E_k^{\rm max}$ determines the upper
integration limit over the electron kinetic energy
in Eq.\ (\ref{ion_t2}). Most calculations in this work are 
carried out with $E_k^{\rm max} = 3.0$\,eV.
The dependence of $\mu_{i0}(E_k)$ on $E_k$
stems from the bound-free dipole integrals ${\rm d_{ks}}(E_k)$ in Eq.\
(\ref{muj0}) and is essentially governed by the overlap of a valence
orbital of CO$_2$ and a distorted plane wave corresponding to the
photoelectron ejected with the kinetic energy $E_k$. This overlap
drops with
increasing $E_k$ leading to narrow photoelectron spectra of these states,
with the measured\cite{BCEKLRTVW96} and calculated\cite{DO79}
electron binding energy widths below 2.0\,eV. Broad photoelectron
kinetic energy distributions exceeding 7.0\,eV, were observed for CO$_2$
initially pre-excited into the Rydberg states near ionization
threshold.\cite{ASSG17} Additionally, electron kinetic energies up to and
exceeding 10\,eV were previously measured in the
ionization of CO$_2$ with XUV APTs.\cite{KWBHC20,KRSGJLLV11} 
In order to assess the role 
of $E_k^{\rm max}$, some calculations for Set 1 were
repeated with $E_k^{\rm max} = 10.0$\,eV. Spectral peaks were reproduced
across the full range of $\omega_{\rm NFT}$. The impact of
$E_k^{\rm max}$ on the peak
intensities was visible only at high NFT frequencies 
above $13\omega_0$. 

\begin{figure}[h!]
\includegraphics[angle=0,scale=0.50]{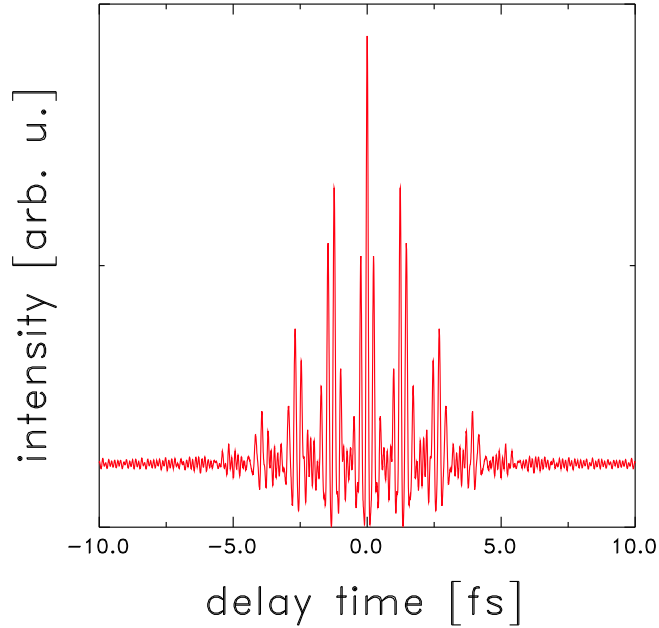}
\vspace{-4.1cm}
\caption{
  Ab initio pump probe NFT signal $I_{\rm C+}(\tau)$
  in a simulation with
  the fundamental frequency of $\omega_0 = $ 1.55\,eV and the amplitude
  Set 1. The optical cycle of the fundamental laser pulse is $T_0 = 2.67$\,fs.
}
\label{nft_time}
\end{figure}

An example of the calculated NFT signal $I_{\rm C+}(\tau)$
in the time domain
is shown in Fig.\ \ref{nft_time} for the pump-probe APTs with the amplitude
Set 1. 
The fundamental harmonic frequency is
$\omega_0 = 1.55$\,eV, close to the experimental value used in
Ref. \onlinecite{OFSNYM14}. The pump probe
time delay $\tau$ in the calculations 
varies between $\pm \tau_{\rm max} = \pm 20$\,fs.
This guarantees a tidy rendering of
the NFT spectra in the frequency domain via Eq.\ (\ref{ion_w}).
The chosen $\tau_{\rm max}$ is about three times larger 
than the experimental value. 

In the NFT signal $I_{\rm C+}(\tau)$ in Fig.\ \ref{nft_time}, most 
of the visible
dynamical effects are concentrated within
$|\tau| \le 7$\,fs, which is about 2---3 times the
optical cycle of the fundamental laser frequency. With growing $\omega_0$,
the signal amplitudes are observed to decrease substantially (not shown in
Fig.\ \ref{nft_time}). The dependence of the NFT signal intensity on the
fundamental laser frequency is considered in more detail in Sect.\
\ref{summary}. 

\begin{figure}[h!]
  \includegraphics[angle=0,scale=0.60]{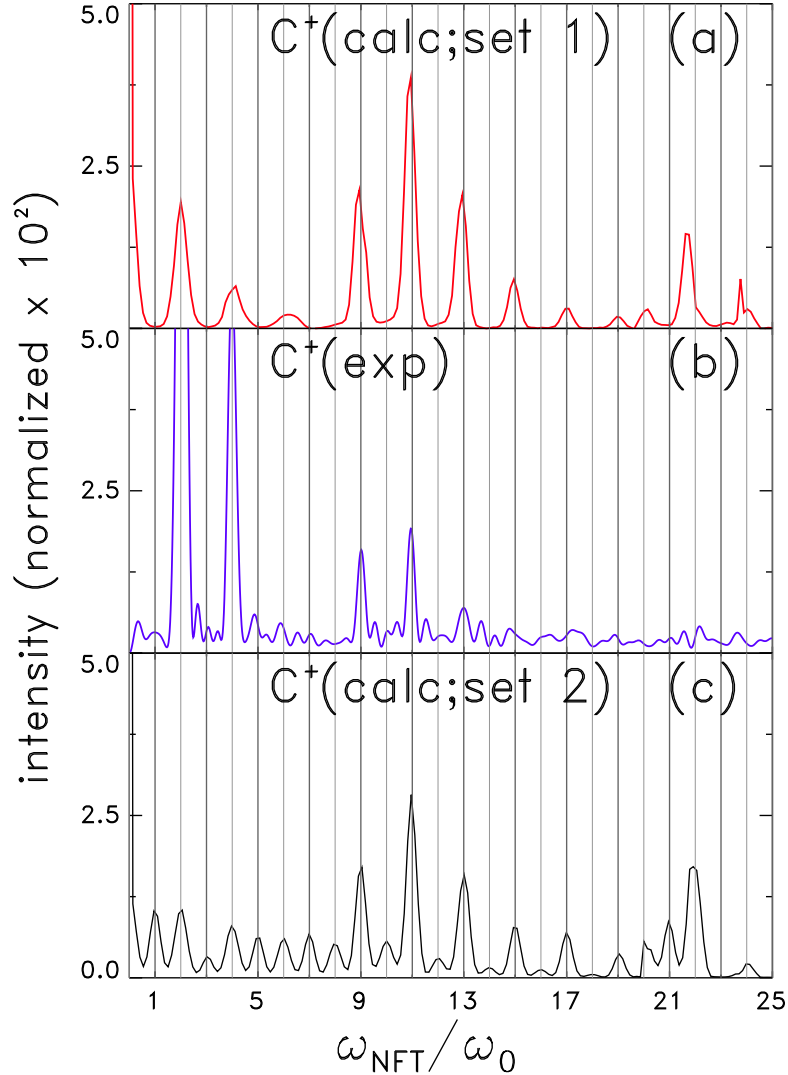}
  \vspace{-0.5cm}
  \caption{
    Ab initio pump probe NFT spectra
    $I_{\rm C+}(\omega_{\rm NFT})$ of CO$_2$
    calculated using APTs with the amplitude Set 1 (a)
    and Set 2 (c).
    The fundamental frequency in the calculations is
    $\omega_0 = $ 1.55\,eV. Shown in (b) is the
    1D NFT spectrum of Ref.\ \onlinecite{OFSNYM14} obtained from the
    experimental 2D
    NFT spectrum for the C$^+$ fragment in their Fig.\
    12(c) by integrating over the fragment kinetic energy. All spectra are
    normalized to the area within the region of 
    $\omega_{\rm NFT} \le 25.0$\,eV. Thick (thin)
    vertical lines indicate odd (even) integer
    harmonic orders $\omega_{\rm NFT}/\omega_0$. 
  }
  \label{nfts_cplus}
\end{figure}

Fourier transform of the signal in Fig.\ \ref{nft_time}
gives the frequency dependent spectrum $I_{\rm C+}(\omega_{\rm NFT})$
depicted in Fig.\ \ref{nfts_cplus}(a). The calculated spectrum consists of
a series of non-overlapping peaks with FWHM of about 0.9\,eV, similar to
the width of the harmonics making up the pump and probe APTs. 
In order to simplify visual comparison
with the spectral content of the incident APTs, the NFT frequency axis
is shown in the units of the fundamental frequency $\omega_0$. 
Two groups of spectral peaks can be identified. The first
group comprises peaks located sufficiently
close to the harmonic orders, either odd,
i.e. present in the incident APTs, or even.
Example is the strong triad at
$\omega_{\rm NFT} \approx 9\omega_0$,
$11\omega_0$, and $13\omega_0$ in panel (a), although the
peaks $9\omega_0$ and $11\omega_0$ slightly but visibly 
deviate from the harmonic position. 
The assignment of the excitation patterns, discussed in Sect.
\ref{assign}, suggests that this triad is due to (a) ionizations into the
the ground electronic state $\tilde{X }^2\Pi_g$ of CO$_2^+$, as well as the
first excited doublet state $A^2\Pi_u$, and (b) dissociations in the
high lying electronic states
belonging to the series $n^2\Pi_g$ and $n^2\Pi_u$.
Previous analyses of the NFT spectra concentrated mainly on the peaks belonging
to this group.\cite{OFSNYM14}
Gradually attenuating contributions of
$15\omega_0$, $17\omega_0$, and $19\omega_0$ can also be recognized. However,
they lie beyond the detection range of the experimental setup limited to less
than $25.0$\,eV or $\omega_{\rm NFT} < 15\omega_0$.\cite{NOTE-NFTS-04}
Note that the
relative intensities of the harmonic peaks in the
NFT spectrum are quite different from those  
in the incident APT [cf. Set 1 in Fig.\ \ref{apt}(a)]. 

Another example of peaks in this group are spectral lines
at even harmonic orders $2\omega_0$ and
$\sim\!4\omega_0$. They are absent in the APTs and have been
previously attributed to the frequency beats between the incident odd
harmonics.\cite{OFSNYM14} In other words, they are seen as difference
frequency peaks stemming from the \lq parent' triad e.g. 
$(13\omega_0 - 11\omega_0)$
for the second harmonic peak and $(13\omega_0 - 9\omega_0)$ for the forth
harmonic peak.

The second group of spectral peaks is formed by the 
peaks located at non-harmonic frequencies. Examples
include the peak between
$6\omega_0$ and $7\omega_0$ and, to a lesser extent, the peaks near
$\sim\!4\omega_0$ and $\sim\!9\omega_0$, 
as well as the high frequency peaks above $20\omega_0$. 

The experimental NFT spectrum $I_{\rm C+}(\omega_{\rm NFT})$
is shown in Fig.\ \ref{nfts_cplus}(b). It
is obtained from the full 2D spectrum published in Ref.\
\onlinecite{OFSNYM14} by integrating over the kinetic energy axis. One 
finds peaks belonging to the same
two groups discussed above. The
triad  near $\omega_{\rm NFT} \approx 9\omega_0$,
$11\omega_0$, and $13\omega_0$ is again
well recognizable. The difference frequency
peaks at $2\omega_0$ and $4\omega_0$ are very strong. This observation
is 
unexpected because the difference peaks turn out to be much stronger
than any \lq parent' harmonic peak in the spectrum. Their
intensities in the normalized spectrum
are six times stronger than in the calculations.
Finally, several non-harmonic peaks
are seen in the experimental spectrum, for example between
$4\omega_0$ and  $6\omega_0$. Note however that the experimental
spectrum is quite congested with a low intensity \lq noisy'
contribution covering the whole experimental NFT frequency range. 

The amplitude Set 1 does not properly describe the experimental APTs of Ref.\ 
\onlinecite{OFSNYM14}, because
many low frequency harmonic components are missing. 
The amplitude Set 2, depicted in Fig.\ \ref{apt} with gray boxes,
is a better model of this experiment. The artificially
amplified amplitude of the fundamental $n=1$  
is immaterial for the case of C$^+$ ion signal
and does not affect the spectrum
$I_{\rm C+}(\omega_{\rm NFT})$ at all. 
The spectrum calculated with Set 2 is shown in
Fig.\ \ref{nfts_cplus}(c). The low frequency harmonics have no impact on 
the spectral peaks of the strong triad 
$9\omega_0 - 11\omega_0 - 13\omega_0$, which are still well recognizable,
similarly to the experimental spectrum. The Set 2 adds a substantial
low intensity component to the spectrum
producing a small spectral peak at every harmonic
order $n$, both even and odd. This feature is also in line with the
experiment. 
Note that the low intensity component is entirely absent in
panel (a), so that the 
\lq noise' in the calculation stems from the harmonic orders $3 \le n \le 7$. 
Nevertheless, the agreement between the calculated and the measured spectra
is at best qualitative, and the relative intensities for many peaks,  
for example $2\omega_0$ and $4\omega_0$, disagree substantially.

\begin{figure}[h!]
  \includegraphics[angle=0,scale=0.60]{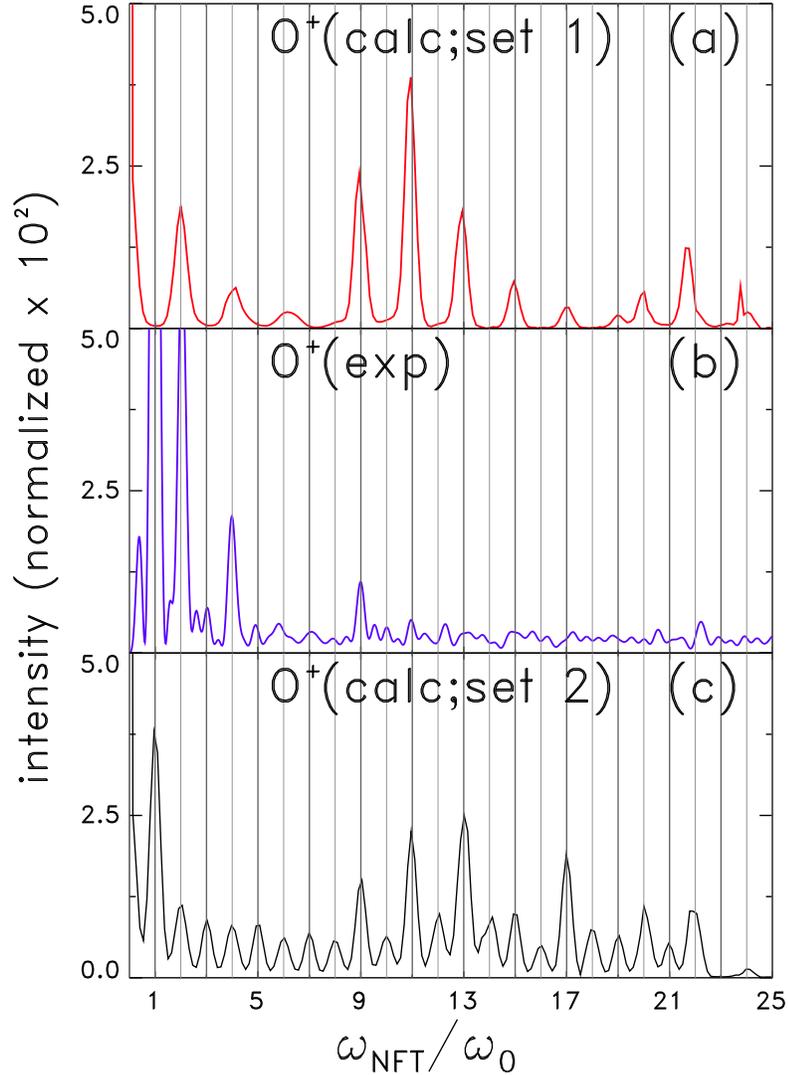}
  \vspace{-0.5cm}
  \caption{
    As in Fig.\ \ref{nfts_cplus}, but for the spectrum
    $I_{\rm O+}(\omega_{\rm NFT})$. The  1D NFT spectrum in panel (b) is
    obtained from the experimental 2D
    NFT spectrum of Ref.\ \onlinecite{OFSNYM14} 
    for the O$^+$ fragment in their Fig.\
    13(c) by integrating over the fragment kinetic energy.
      }
  \label{nfts_oplus}
\end{figure}

Lowering the appearance threshold from 30.0\,eV to 19.0\,eV, one can use
the same ladder ab initio electronic states to evaluate the NFT spectrum
$I_{\rm O+}(\omega_{\rm NFT})$ for the fragment ion O$^+$. The results are
summarized in Fig.\ \ref{nfts_oplus} for the amplitude Set 1 in panel (a) and
Set 2 in panel (c). The experimental spectrum is shown in panel (b).

The experimental spectra recorded for O$^+$ and C$^+$ are similar in
several respects. For example, the triad near 
$9\omega_0 - 11\omega_0 - 13\omega_0$ is still visible in
$I_{\rm O+}(\omega_{\rm NFT})$ (although the
peaks are shifted more clearly
to non-harmonic positions), and the low energy peaks
at $2\omega_0$ and $4\omega_0$ are strong. However, the experimental  spectrum 
$I_{\rm O+}(\omega_{\rm NFT})$ is dominated by the very intense
peak $n=1$ of
the fundamental frequency $\omega_0$. Additionally, there are more
non-harmonic peaks in the O$^+$ spectrum than
in the C$^+$ spectrum, and their intensity is higher. 

The amplitude Set 1 is clearly incompatible with experiment: 
The spectrum in Fig.\ \ref{nfts_oplus}(a)
is almost the same as in  Fig.\ \ref{nfts_cplus}(a)
and is insensitive to a shift in the appearance
threshold. Indeed,
the amplitude Set 1 comprises harmonics with energies above 14\,eV, so that
the high lying states of CO$_2^+$ are primarily 
populated along the photochemical pathway of Eq.\ (\ref{chemeq1}). In the
adopted model, these states contribute to both O$^+$ and C$^+$ signals.

The O$^+$ spectrum calculated with Set 2 and shown 
Fig.\ \ref{nfts_oplus}(c) is different and has 
a pronounced peak at $\omega_{\rm NFT}/\omega_0=1$. In the C$^+$ spectrum,
calculated
with the same Set 2, this peak is much weaker.  
Low order harmonics in Set 2 induce 
excitations between neighboring electronic
states of CO$_2^+$ and this directly influences the NFT spectrum.
For example, the
spacing between the states $1^2\Sigma_u^+$ and
$1^2\Sigma_g^+$ is 1.26\,eV (see Table\ \ref{abinitio_en}) and is  
nearly resonant with $\omega_0$.
Still, a single resonant electronic transition in CO$_2^+$ alone
is not sufficient  to make the peak $n= 1$ intense: 
In the calculations, the large
amplitude  $a_1$ of the fundamental
in the incident APT is also a pre-requisite; the chosen amplitude $a_1$
exceeds the value inferred from Ref.\
\onlinecite{OFSNYM14}, by about a factor of 20. It is
therefore likely that the very strong peak at $\omega_0$
in the experimental spectrum has a different origin 
missing in the present calculations. One possibility
are higher order multiphoton excitations.\cite{OFSNYM14} The other are
contributions
from the photochemical pathway of Eq.\ (\ref{chemeq2}), in which CO$_2$ is
photoexcited into the vicinity of the ionization threshold and then ionized
with an infrared photon. Note that the reversed order of absorptions would be
unfeasible because CO$_2$ is essentially transparent around 1.55\,eV. 
Finally, there could be yet another explanation for the large amplitude $a_1$
of the fundamental in the incident APT
needed to reconcile the experiment and theory.
In the traditional high harmonic
generation setups, the spectral filters and the reflectiveness of XUV mirrors
are expected to attenuate the fundamental frequency by a factor of
$10^{-4}$ to $10^{-5}$.
One can therefore reasonably expect that sufficient amount of the fundamental
may reach the molecular system to affect its time-domain response and
NFT spectra. 

\subsection{Assignments of the calculated spectra: Near-harmonic and
non-harmonic peaks}
\label{assign}

The comparison between the experimental and the ab initio spectra, discussed
in the previous section, offers several insights into the 
relation between the NFT spectrum and the composition
of the incident APTs. We turn now to 
the assignment of the spectral peaks, which allows one to relate
the NFT spectra to the properties
of the electronic states excited in the pump probe experiment and to expose the
origin of near-harmonic and non-harmonic spectral peaks. 
Our discussion is based 
on the calculations using the amplitude Set 1
and $\omega_0 = 1.55$\,eV: The spectrum 
$I_{\rm C+}(\omega_{\rm NFT})$, calculated with these settings, shares
many principal features with the experimental spectrum. 

\begin{figure}[]
  \includegraphics[angle=0,scale=0.70]{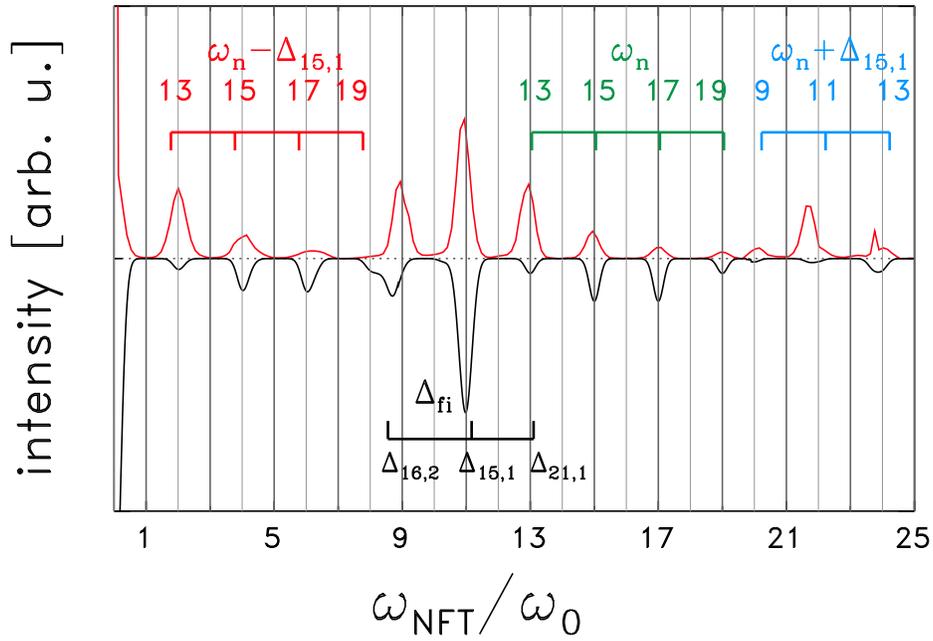}
  \vspace{0.5cm}
  \caption{
    Comparison of the numerically exact quantum mechanical
    NFT spectrum
    (red line) with the approximation of 
    Eq. (\ref{nfts-approx}) (black line).
    All calculations are performed for $I_{\rm C+}(\omega_{\rm NFT})$ 
    using APTs with the amplitude Set 1. 
    The fundamental frequency is 
    $\omega_0 = $ 1.55\,eV.     Combs show the assignments in
    terms of the harmonic frequencies $\omega_n$ and the energy differences
    $\Delta_{fi}$ between the ionization and the final dissociation states. 
    States contributing to the assigned peaks are
    numbered as in Table \ref{abinitio_en}: $1 = 1^2\Pi_g$, 
    $2 = 1^2\Pi_u$, $15 = 8^2\Pi_u$, $16 = 10^2\Pi_g$, and $21 = n^2\Pi_u$.
    Thick (thin)
    vertical lines indicate odd (even) integer
    harmonic orders $\omega_{\rm NFT}/\omega_0$.      }
  \label{nfts_assign}
\end{figure}

Appendix \ref{appb} sets the stage for the discussion and
summarizes the derivations of analytic expressions for
$I_{\rm ion}(\omega_{\rm NFT})$  which can be used to guide the
spectral assignment. Specifically, 
Eqs.\ (\ref{nfts-approx}) and (\ref{nfts-approx2}) 
provide explicit dependence of the spectral intensity
on the NFT frequency $\omega_{\rm NFT}$, on the transition frequencies
between electronic states, $\Delta_{fi} = \epsilon_{0f} - \epsilon_{0i} >0$,
and on the harmonic
frequencies $\omega_n$ of the APT components. They are derived
using the approximation akin to the temporarily non-overlapping pump
and probe pulses, familiar in the context of ultrafast pump probe
spectroscopies.\cite{SSD89,SD91A,L95A,WEIHNACHTPEARSON19,NOTE-NFTS-02} 
The accuracy of the approximation of Eq.\ (\ref{nfts-approx})
is illustrated in Fig.\ \ref{nfts_assign}. Most of the
approximate spectral lines (black) are located at the positions of the
numerically exact peaks (red). The intensities,
which are more sensitive to the coherent two photon effects, are less accurate
but the overall spectrum is reasonably reproduced.

The NFT spectrum derived in Appendix \ref{appb} is a 
superposition of the ionizations of
the parent molecule and subsequent excitations into the final dissociative
state. It is therefore natural to label
the NFT peaks using electronic pairs $(i,f)$ comprising
the ionized state(s) $i$ of CO$_2^+$, 
reached in the reaction CO$_2 \rightarrow$\,CO$_2^+(i)$, and the final
dissociative state(s) $f$ of CO$_2^+(f)$
contributing to the observed C$^+$ or O$^+$ signals.
The second set of labels are the harmonic assignments indicating
the photon frequencies $\omega_n$
promoting specific transitions leading to the ionization and
dissociation. In practice, we scan the sums in Eq.\ (\ref{nfts-approx}) for
the terms making the largest contribution at a specified $\omega_{\rm NFT}$.

The assignments of the calculated spectrum 
in terms of these attributes 
are summarized in Fig.\ \ref{nfts_assign}.
The intermediate states in the ionization step
include the ground electronic state
of CO$_2^+$,  $\tilde{X}^2\Pi_g$ as well as the first excited doublet state
$\tilde{A}^2\Pi_u$. The final
electronic states contributing to the fragment ion signal belong to the
series $n^2\Pi_g$ and $n^2\Pi_u$. 

In fact, two distinct assignment schemes emerge for the spectral peaks
--- and they are best illustrated using 
Eq.\ (\ref{nfts-approx2}) which is valid for the specific case of
only one intermediate ionized state $i$ contributing to the dissociation
via the final state $f$. The spectrum  $I_{\rm C+}(\omega_{\rm NFT})$ 
in this case consists of two additive contributions. The first is given by
\begin{align}
\label{part1}
\notag & 
\left(\mu_{fi}\mu_{i0}\right)^2
\sum_{n_1,n_2,n_3,n_4}
  a_{n_1}a_{n_2}a_{n_3}a_{n_4} 
{\cal A}_{n_2,n_4}L_{n_1}(\Delta_{fi})L_{n_3}(\Delta_{fi}) \\
\times & \Big\{
4\delta_\tau(\omega_{\rm NFT})
+ 2\delta_\tau(\omega_{\rm NFT}- \Delta_{fi})
+ 2\delta_\tau(\omega_{\rm NFT}+ \Delta_{fi})
\Big\} \, .
\end{align}
Here $L_{n_i}(\Omega) = e^{-(\Omega - \omega_{n_i})^2T^2/4P}$ are
the Gaussian Fourier images of the time envelope
function $L_{n_i}(t,\tau)$  of the APTs defined in 
Eqs.\ (\ref{etot})---(\ref{eapt2}). The transition is mediated by four
photons with harmonic frequencies $n_1$, $n_2$, $n_3$, and $n_4$. 
The resulting NFT spectral
peaks are located at $\omega_{\rm NFT} = 0$ (the zero peak
which we do not consider) and 
at the energy differences between the intermediate
ionized and the final dissociative states,
$\omega_{\rm NFT} = \pm \Delta_{fi}$. They are denoted $\delta_\tau$ 
because their
width is only controlled by the maximum
delay time in the time signal, and can potentially
be made narrow (delta-function like) by increasing $\tau_{\rm max}$. Note that
these spectral peaks can appear at non-harmonic frequencies if the energy
$\Delta_{fi}$ is off resonance. The calculation with the Eq.\
(\ref{nfts-approx}) recognizes two such spectral lines in Fig.\ 
\ref{nfts_assign}: One near  $\omega_{\rm NFT} \approx 9\omega_0$ and the
other at
$11\omega_0$. They belong to the strong central triad discussed in the previous
section. We also use this assignment for the third member of the triad,
$\omega_{\rm NFT} \approx 13\omega_{0}$. The peaks of the triad exemplify
the appearance of near harmonic and non-harmonic peaks in the spectrum. 
The pair of electronic states $(i=\tilde{X}^2\Pi_g,\,f=8^2\Pi_u)$
has, according to Table\ \ref{abinitio_en},
$\Delta_{fi} = 17.08$\,eV which is nearly resonant with the harmonic frequency
for
$n = 11$ (17.05\,eV). It gives rise to an intense line at this harmonic order.
The pair of states $(i=\tilde{A}^2\Pi_u,\,f=10^2\Pi_g)$ is spaced by
13.49\,eV, which is off resonance with respect to
the harmonic frequency $\omega_9 = 13.95$\,eV. The detuning
affects the intensity
of this off-resonance non-harmonic line. Indeed, the Gaussian prefactors 
$L_{n_1}(\Delta_{fi})L_{n_3}(\Delta_{fi})$ suppress large deviations of
$\Delta_{fi}$ from integers $n_1 = n_3 = 9$. However, the Gaussians are
almost 1.0\,eV broad and --- additionally --- $\tau_{\rm max}$ is large
(20\,fs) in this
calculation, so that the intensity of this non-harmonic line
becomes appreciable. Finally,
the peak near $\omega_{\rm NFT} \approx 13\omega_{0}$
can be plausibly assigned to the pair of states
$(i=\tilde{X}^2\Pi_g,\,f=n^2\Pi_u)$ which are in resonance with the
harmonic frequency for $n = 13$ (20.15\,eV); the offset is merely 0.1\,eV, and
the third member of the triad is perceived as a harmonic peak. 

The second contribution to the NFT spectrum in Eq.\
(\ref{nfts-approx2}) has a different form:
\begin{align}
\label{part2}
\notag &  \left(\mu_{fi}\mu_{i0}\right)^2
\sum_{n_1,n_2,n_3,n_4}
a_{n_1}a_{n_2}a_{n_3}a_{n_4} \
L_{n_1}(\Delta_{fi})L_{n_3}(\Delta_{fi}) \\
\notag \times & \Big\{
  2L_{n_2}(\omega_{\rm NFT})L_{n_4}(\omega_{\rm NFT}) 
 +
  L_{n_2}(\omega_{\rm NFT}-\Delta_{fi})
  L_{n_4}(\omega_{\rm NFT}-\Delta_{fi}) \\
 & + L_{n_2}(\omega_{\rm NFT}+\Delta_{fi})
  L_{n_4}(\omega_{\rm NFT}+\Delta_{fi})  
  \Big\}
\end{align}
There are three sets of peaks associated with this contribution:
One set is located
at the original harmonic
frequencies, $\omega_{NFT} \approx
\omega_{n_2}$ or $\omega_{n_4}$. In Fig.\ \ref{nfts_assign}, they are shown
with the green comb. These peaks are  
locked on the spectral composition of the incident APTs and have
little to no dependence on the ionic energy levels.
The third member of the
strong triad at $13\omega_0$ contains a low intensity contribution
of this type. In the second set,  
these harmonic peaks are shifted to
lower energies by the energy difference $\Delta_{fi}$. 
In Fig.\ \ref{nfts_assign}, they are shown
with the red comb, and the electronic states are again 
$i = \tilde{X}^2\Pi_g$ and $f = 8^2\Pi_u$.
Because the energy spacing $\Delta_{fi}$
is in resonance with $\omega_{11}$, the shifted peaks appear at
even harmonics between $n=2\omega_0$ and $n=8\omega_0$ and are 
interpreted as difference
peaks in the NFT spectra. According to Eq.\ (\ref{part2}),
their intensity is 
one half the intensity of the unshifted harmonic peaks. 
In the third set, 
the harmonic peaks are shifted to higher energies
by the same amount $\Delta_{fi}$. They are located at high NFT frequencies
and marked with the blue comb. In the simple approximation which we are
using, 
their information content is the same as in 
spectral peaks shifted to low $\omega_{\rm NFT}$. In the present calculation,
the peaks of all three sets, even the shifted ones,
appear near the integer harmonic orders,
either even or odd. This is because the electronic states involved in the
optical transitions within the ion are in resonance with one harmonic
frequency. However, 
each NFT peak in Eq.\ (\ref{part2})
is a product of four Gaussian factors $L_{n_i}$, and each Gaussian
has a width of the
order of $\omega_0/2$. Thus, peaks located at non-harmonic frequencies can
in principle be expected for the shifted lines. In fact,
deviations from integer harmonic orders can be seen for the
peaks at $\sim 4\omega_0$ and $\sim 6\omega_0$, as well as $\sim 24\omega_0$. 
These deviations can be related to the specific electronic transitions
in the molecular ion. 

\begin{figure}[h!]
  \includegraphics[angle=0,scale=0.70]{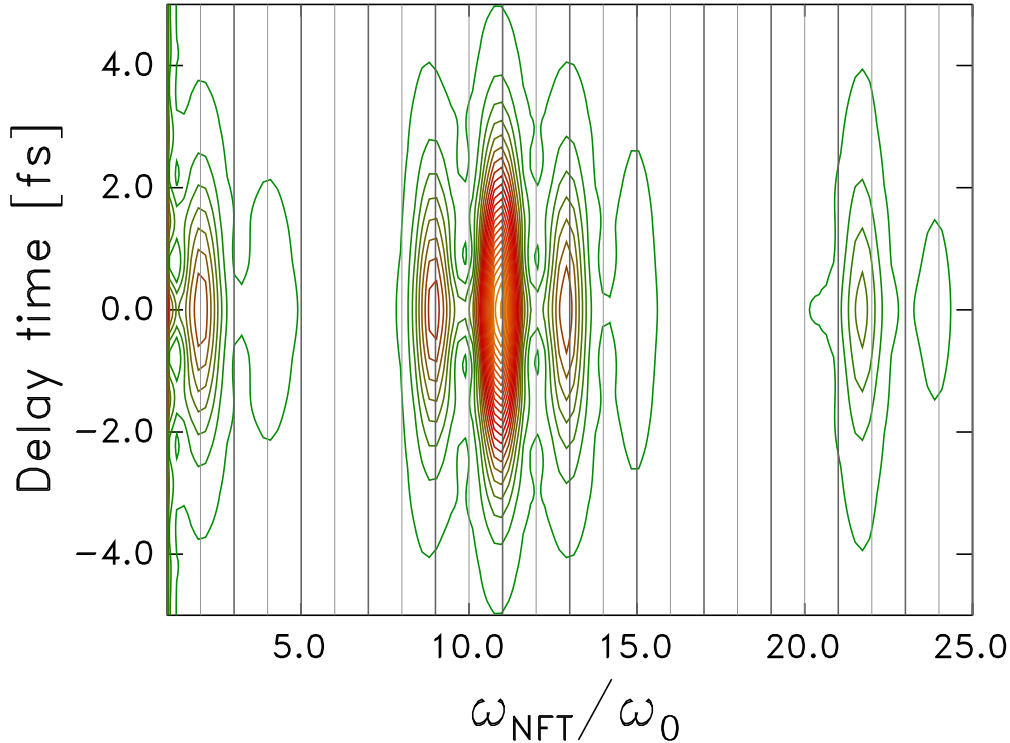}
  \vspace{0.0cm}
  \caption{ The spectrogram $S_{\rm C+}(\omega_{\rm NFT},t)$
    as defined in Eq.\ (\ref{vibro}), with $\omega_{\rm NFT}$ plotted in the
    units of $\omega_0$. Red contours correspond to the maxima.
    The calculations are performed 
    using APTs with the amplitude Set 1. 
    The fundamental frequency in the calculations is
    $\omega_0 = $ 1.55\,eV. Vertical lines mark even and odd
    harmonic orders. 
      }
  \label{nfts_vibro}
\end{figure}

Spectral peaks at different NFT frequencies can be further
characterized by the ranges of pump probe delay times, at which they
are formed. The relation between $\omega_{\rm NFT}$ and $\tau$ is 
visualized using the so-called spectrogram 
\begin{equation}
  \label{vibro}
  S_{\rm C+}(\omega_{\rm NFT},t) =
  \int I_{\rm C+}(\tau)  h(\tau-t) e^{i\omega_{\rm NFT}\tau}\,d\tau \, ,
\end{equation}
which is a moving window Fourier transform of the pump probe signal.
Here $h(\tau) = \exp(-\tau^2/2\tau_0^2)$ is the Gaussian window function.
In the spectrogram
$S_{\rm C+}(\omega_{\rm NFT},t)$, shown in Fig.\ \ref{nfts_vibro},
the delay time and the NFT frequency domains are
represented in the same plot at
the expense of the resolution which is smeared along both
axes:\cite{JK89,HHG95,VVS96} The time resolution $\tau_0$
in the plot is 0.9\,fs
and corresponds to the frequency resolution of
$\tau_0^{-1} = 0.80$\,eV, about half the fundamental frequency $\omega_0$. 

All major spectral peaks can be clearly recognized in the spectrogram,
and they acquire an additional --- time delay --- dimension.
The spectrogram provides a different perspective on the assignment and
helps to identify spectral features observed over the same ranges of the delay
times. For example, the peaks in the strong central triad 
$9\omega_0 - 11\omega_0 - 13\omega_0$ have similar spectrograms and 
collect their intensities over a broad
time delay range from -5.0\,fs to +5.0\,fs. This is an additional reason to
assign them using one common assignment scheme as done in Fig.\
\ref{nfts_assign}.
The peaks at $2\omega_0$ and at $22\omega_0$ extend over
a similarly broad time delay range --- they both belong follow the assignment
in terms of $\omega_n\pm \Delta_{fi}$ and share the electronic states $f$ and
$i$ with the central peak of the triad. For the weaker peaks in the spectrum,
the relationships within spectral groups are less straightforward. For
example, the peaks at $\sim4\omega_0$, $15\omega_0$, and $\sim 24\omega_0$
are due to short delay times $|\tau| \le 2.0$\,fs. The spectrogram
indicates that they might belong to a separate assignment group which we were
not able to identify yet. 
\begin{figure}[h!]
  \includegraphics[angle=0,scale=0.70]{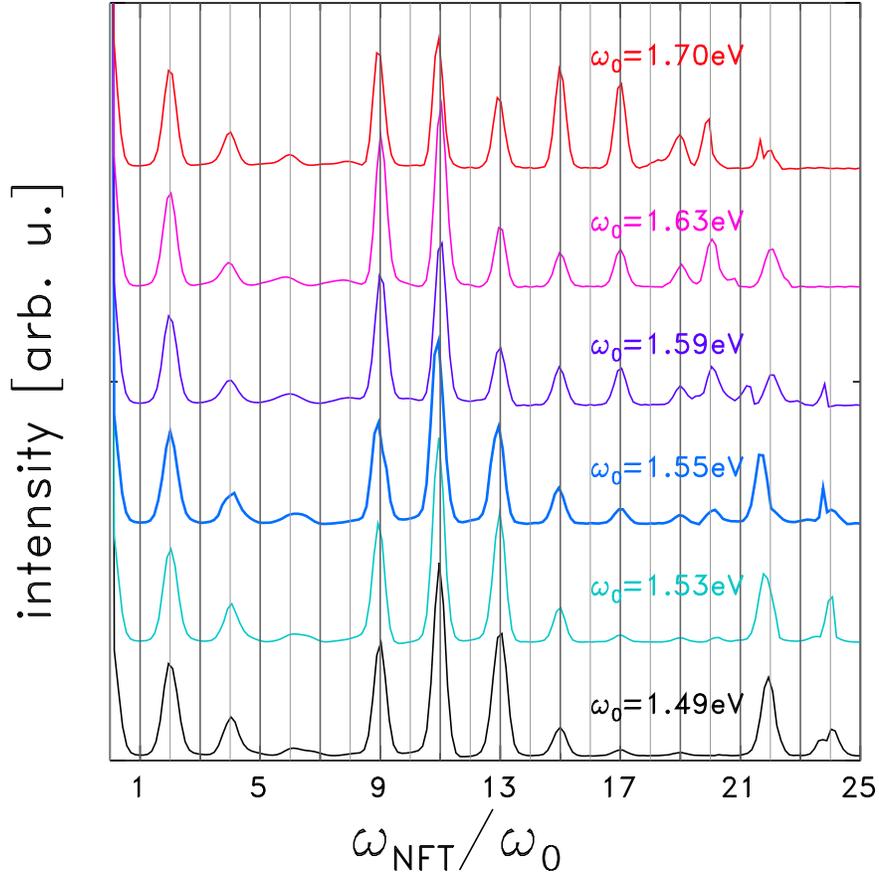}
  \vspace{-0.5cm}
  \caption{
    Ab initio pump probe NFT spectra
    $I_{\rm C+}(\omega_{\rm NFT})$ of CO$_2$ for a set of fundamental
    harmonic frequencies $\omega_0$.
    The incident APTs are constructed using
    the  amplitude Set 1. All spectra are
    normalized to the area within the region of 
    $\omega_{\rm NFT} \le 25.0$\,eV. Thick (thin)
    vertical lines indicate odd (even) integer
    values of the ratio $\omega_{\rm NFT}/\omega_0$.
  }
  \label{nfts_over}
\end{figure}

In summary, the developed approach 
provides a reasonable description of NFT spectra of CO$_2$, shows their
relation to the experiment, 
and rationalizes the spectral assignments 
demonstrating the origin of near harmonic and
non-harmonic peaks in the spectra.

\section{Conclusions and outlook}
\label{summary}

In this paper, we outlined a quantum mechanical approach to modeling
attosecond NFT spectra of CO$_2$ ionizing to CO$_2^+$.
The approach combines perturbation theory
for the molecule-light interaction with ab initio calculations of the
electronic energy levels of CO$_2^+$. The ab initio calculations are performed
using accurate MRCI method accounting for the electron correlation, but
in this work they are limited to the Franck-Condon zone --- so that the
resulting ladder of electronic levels of CO$_2^+$ represents a \lq toy
model' of the molecular ion. The main
results can be summarized as follows: 
\begin{enumerate}
\item One-dimensional NFT spectra of CO$_2$ are calculated for two different
  incident APTs and compared with the available experimental spectra. Several
  features of the experiment, including the positions and intensity patterns
  of the main harmonic spectral peaks, the presence of the difference
  frequency peaks, and the origin of the low intensity contribution at all
  harmonic orders, are reproduced and/or explained.
\item  The calculations give an overview of the NFT spectrum outside the
  experimental frequency window, for $\omega_{\rm NFT} \ge 25.0$\,eV, where
  strong excitations can be expected.  
\item A set of approximate
  analytical expressions for the spectral intensity is derived, demonstrating
  how the NFT spectral intensity depends 
  on the frequency $\omega_{\rm NFT}$,
  the harmonic  frequencies $\omega_n$ of the APT components, and
  the transition frequencies
  between electronic states. These expressions guide the 
  assignment of the spectral peaks. 
\item The calculated NFT spectra are assigned in terms of the
  participating electronic states and the harmonic photon frequencies.
  The  assignment
  demonstrates which details of the electronic structure of the
  CO$_2^+$ are captured in the NFT experiments. Using spectrograms, the
  NFT peaks can be additionally attributed to specific ranges of the
  pump probe delay times. 
\item It is shown
  that spectral peaks at non-harmonic frequencies can be
  expected, especially if APTs with spectral bandwidths of about
  1\,eV are used and photoreactions are limited to single ionizations.
  Non-harmonic NFT spectral peaks carry additional
  information on the electronic
  states mediating ionization of the parent molecule and dissociation of the
  molecular ion.
\end{enumerate}

The main goal of this work was to set up a framework, within which
NFT spectra can be calculated and analyzed, and to test this
framework for CO$_2$. The tests demonstrate that the framework works with
reasonable precision and can be used for semi-quantitative predictions,
even though CO$_2^+$ is described using an
ab initio toy model. The simplicity of the ab initio calculations limited
to a single Franck-Condon geometry makes this framework highly scalable:
It can be applied to polyatomic molecules and ions and one can easily scan
through various control parameters of the experiment, such as
fundamental laser frequency $\omega_0$, spectral composition of the incident
APTs, or the maximum time delay $\tau_{\rm max}$. 

An example is provided in Fig.\ \ref{nfts_over} which shows a series of
NFT spectra of CO$_2$ calculated using 
$\omega_0$ varying from 1.49\,eV (bottom spectrum) to
1.73\,eV (top spectrum). The main triad $9\omega_0 - 11\omega_0 - 13\omega_0$
is present in all spectra, the intensity distribution between the
peaks changes slightly, and the peak at $13\omega_0$ slowly
attenuates as $\omega_0$ grows. The peaks at higher integer harmonic
orders, present in the incident APTs, clearly become stronger with
increasing $\omega_0$, as more directly ionizing states
become energetically accessible from the ground state of the parent CO$_2$. 
The high frequency end of the NFT spectrum above $19\omega_0$, which
stems from the short time delays $|\tau| \le 2$\,fs and in which the
concerted two photon effects are pronounced, demonstrates a strong 
dependence on $\omega_0$, with multiple non-harmonic contributions.
In contrast, the low frequency region of the difference frequency peaks is
stable, and the peaks $2\omega_0$ and $\sim 4\omega_0$ 
change neither position nor intensity. Exception is the weaker 
non-harmonic difference peak which moves between
$6\omega_0$ and $7\omega_0$.  Figure \ref{nfts_over}, together with the
approximate assignment schemes illustrated in Fig.\ \ref{nfts_assign}, 
could in principle be used as a starting point 
for spectral inversion analysis, in which the electronic states of
the dissociating ion are reconstructed from the NFT spectra.

It is highly desirable
to extend the developed framework and to replace
the toy model based on the
electronic state ladder with realistic and interacting
potential energy surfaces. This extension, which will be  presented in
a separate publication, removes two main limitations of the
discussed approach. First, it
allows calculations of two-dimensional NFT spectra as functions of both
fragment kinetic energy and $\omega_{\rm NFT}$. 
Appendix \ref{appa} provides the necessary
formal expressions. Second,  vibronic interactions between
electronic states 
become naturally incorporated into the model so that one can analyze the 
sensitivity of the NFT spectra 
to non-adiabatic interactions in the parent molecule and in the
molecular ion. The time scales and
the atomistic mechanisms of ultrafast photoreactions mediated by the vibronic
interactions 
are actively explored\cite{CIBOOK12,CSWYDC16} across a
wide range of applications from advanced energy materials\cite{LANZANI06} to
photoprotective mechanisms in biochromophores.\cite{FA93,MBRD08} 
In CO$_2$ and CO$_2^+$, non-adiabatic interactions
affect not only the absorption and ionization profiles but also
the lifetimes of electronically excited species, and the
dissociation mechanisms can be controlled by vibronic as well as
spin-orbit (i.e. relativistic) interactions.

\appendix
\section{Two-dimensional NFT signals}
\label{appa}

After the interaction with two APTs, the excited ion
$\left({\rm CO}_2^+\right)^{\star\star}$ and the
photoelectron are in the state 
$\Psi_I(t|E_k,\tau)$. We first consider one 
electronic state $|f\rangle$ of CO$_2^+$ which
dissociates into an arrangement channel with the detected fragment, and
restrict the description to 2-body
arrangement channels, as indicated in  Eq.\ (\ref{chemeq3}). In this
case, one of the dissociation fragments is diatomic and its internal state
is characterized by rovibrational quantum numbers, which we collectively
denote $n_f$, and by the internal energy $\epsilon_{\rm int}(n_f)$.
Quantum numbers $n_f$
label individual dissociation channels of
$\left({\rm CO}_2^+\right)^{\star\star}$ in the state $|f\rangle$
in the considered arrangement channel. 
The corresponding dissociation
threshold is $D_f$. Thresholds relevant for the
NFT experiments on CO$_2$ are illustrated in Fig.\ \ref{fig_energies}. 

Suppose that the final energy of
$\left({\rm CO}_2^+\right)^{\star\star}$ after the absorption of two photons 
has a value of $\epsilon_f$
lying above $D_f$. In the dissociation channel, this energy is shared between
the (center of mass)
recoil kinetic energy $\epsilon_{\rm kin}$ and the internal excitation: 
\begin{equation}
\label{efrag}
\epsilon_f = \epsilon_{\rm kin} + \epsilon_{\rm int}(n_f) > D_f \, .
\end{equation}
The final rovibrational distribution and the final kinetic energy
distribution are complementary and can be recalculated from one another
using the energy conservation. 

The state of the dissociating fragments is a linear combination of the
scattering states\cite{PERELOMOV98} $\psi^-_{f,k_f,n_f}$, corresponding to
the total energy $\epsilon_f$.  The probability
amplitude for the fragment to be in a state with energy $\epsilon_f$ and
wave vector $k_f$ (and the kinetic energy
$\epsilon_{\rm kin}(k_f) = k_f^2/2\mu$), while the diatomic fragment is in the
internal state $n_f$, is given by the matrix element 
\begin{equation}
  \label{photomatrx}
  \gamma(\epsilon,k_f,n_f|E_k,\tau) =
  \langle \psi^-_{f,k_f,n_f}| \langle \phi_f^+\psi^e_k
  |\Psi_I(t\rightarrow\infty|E_k,\tau)\rangle\, .
\end{equation}
This expression is akin to the photodissociation
matrix element\cite{BK04} which contains the dynamical information on the
dissociation process.\cite{PG18A,PG18B} The above expression is applied to the
ionizing system, and the projection on $\langle \phi_f^+\psi^e_k|$
additionally specifies that 
the ejected photoelectron leaves with the energy $E_k$.

The partial cross section to produce the detected fragment ion in
a given dissociation channel $n_f$ describes the rovibrational state
distribution 
--- and equivalently the kinetic energy distribution --- for the fixed
energy $\epsilon_f$. It is given by the square of the matrix element
$\int\,d\,E_k|\gamma(\epsilon,k_f,n_f|E_k,\tau)|^2$
integrated over all possible photoelectron
energies. However, this kinetic energy
distribution is not accessible in the multiphoton NFT spectroscopic
experiment. The 2D NFT signal as measured, for example, in
Ref.\ \onlinecite{OFSNYM14}, is additionally summed over all possible final
energies $\epsilon_f$ and all final electronic states $|f\rangle$
contributing to the detected dissociation fragment:
\begin{equation}
  \label{i_et}
  I_{\rm ion}^{2D}(\epsilon_{\rm kin},\tau) = \sum_f
\int_{D_f}^\infty d\,\epsilon_f \int_0^\infty d E_k\,
|\gamma(\epsilon,k_f,n_f|E_k,\tau)|^2 \, .
\end{equation}
Evaluation of the photodissociation matrix element
$\gamma(\epsilon,k_f,n_f|E_k,\tau)$ requires numerical 
solution of the Schr\"odinger
equation with the (generally three-dimensional) potential energy
surfaces of the dissociative electronic states of CO$_2^+$. While the nuclear
dynamics for the photofragment distributions 
can be efficiently calculated using iterative
methods,\cite{PG14,PG15,PG16B} construction of the multidimensional 
potential energy surfaces of many densely spaced 
(and possibly interacting) electronic states
in the energy range illustrated in Fig.\ \ref{fig_energies}
is a true challenge.\cite{GQZSCH07,G13A} 

Sum over all channel quantum numbers $n_f$ effectively brings about summation
over all fragment kinetic energies and gives the 1D NFT signal, i.e.
the total ion yield:
\begin{equation}
  \label{i_t}
  I_{\rm ion}^{1D}(\tau) = \sum_f \sum_{n_f}
\int_{D_f}^\infty d\,\epsilon \int_0^\infty d E_k\,
|\gamma(\epsilon,k_f,n_f|E_k,\tau)|^2 \, .
\end{equation}
The scattering basis states are complete, and this expression
can be rewritten in terms of the populations of the final electronic
states integrated over the photoelectron kinetic energy: 
\begin{equation}
  \label{i_t2}
  I_{\rm ion}^{1D}(\tau) = \sum_f \int_0^\infty d E_k\,
  \langle\Psi_I^c(t\rightarrow\infty|E_k,\tau)|
  \phi_f^+\psi^e_k\rangle\langle \phi_f^+\psi^e_k|
  \Psi_I^c(t\rightarrow\infty|E_k,\tau)\rangle
\, .
\end{equation}
This expression is practically identical to the total ion yield defined
in Eqs.\ (\ref{pop0}) and (\ref{ion_t}) of the main text. The appearance
threshold $A_{\rm ion}$ for the detected ion fragment
corresponds to the lowest energy $\epsilon_f$ (and the lowest state
$|f\rangle$) for which the amplitude $\gamma(\epsilon,k_f,n_f|E_k,\tau)$
does not vanish. The subscript {\it c} on the wave function 
in Eq.\ (\ref{i_t2}) is a remainder that only projections on the
continuum states $\psi^-_{f,k_f,n_f}$ are considered in each state
$|f\rangle$. 


\section{Analytic expressions for 1D NFT signals}
\label{appb}

The quantum mechanical expressions for the NFT  signals
$I_{\rm ion}(\tau)$ and $I_{\rm ion}(\omega_{\rm NFT})$, as described in
Sect.\ \ref{theo1_pt} [see  
Eqs.\ (\ref{ion_t}), (\ref{ion_w}), and (\ref{cvj})---(\ref{ion_t2})],
are well suited for numerical calculations. One can cast them into a
different form more appropriate for the 
analysis and assignment of the spectral peaks. 
This appendix summarizes 
analytic expressions for
$I_{\rm ion}(\omega_{\rm NFT})$ which help to rationalize  the
NFT spectra. The implications for the spectral
assignments
are discussed in Sect.\ \ref{assign}. 

For simplicity, we consider only the photoreaction pathway
in which the parent molecule is ionized
already with the first photon. The second order
pump probe amplitude $c_j(v_j;E_k,\tau)$
is given by the Eq.\ (\ref{cvj}).
Changing the variables to $y = t_1 - t_0$ in the inner integral
with $y$ running from 0 to $\infty$, using
the Fourier transform of the APT electric field, ${\cal E}(\Omega;\tau)$, and
invoking the convolution theorem, the amplitude $c_j(v_j;E_k,\tau)$ can
be re-written as
\begin{eqnarray}
c_j(v_j;E_k,\tau) & = &  \frac{1}{i^2}\sum_{i\ne j}\sum_{vi}\mu_{ji}
\mu_{i0}(E_k) \nonumber \\
&\times& \int_{-\infty}^\infty \frac{d\Omega}{2\pi}
   {\cal E}(\epsilon_{vj} - \Omega;\tau)
   \frac{i\langle v_j | v_{i}\rangle\langle v_{i}|0\rangle}
        {\Omega - \epsilon_{vi} +i0}
{\cal E}(\Omega + E_k - \epsilon_0;\tau) \, .
\label{cvj-ft}
\end{eqnarray}
This (still exact)
pump probe amplitude is given by the convolution of two Fourier images of
the laser electric field and the pole
factor $i\left(\Omega - \epsilon_{vi} +i0\right)^{-1}$ corresponding to
the Fourier transform of $e^{-i\epsilon_{vi}y}$ times 
the Heaviside step function $\Theta(y)$.\cite{SSD89}
The Fourier transform of the laser field is given by 
\begin{equation}
\label{eapt1-ft}
{\cal E}(\Omega;\tau) = \sum_{n}{}^{'} a_n L_n(\Omega)
\left(1 + e^{i\Omega\tau}\right)
\, ,
\end{equation}
with $L_n(\Omega)$ being the Fourier image of the time envelope
function $L_n(t,\tau)$  defined through the
Eqs.\ (\ref{etot}), (\ref{eapt1}), and (\ref{eapt2}). 

The integral over $\Omega$ has a suggestive form for the use of contour
integration in the complex plane and the residue theorem. 
The success of this approach depends on the actual shape of the function 
$L_n(\Omega)$. In the main text, computations are
performed using the Gaussian time 
envelope $L_n(t,\tau)$. The 
function $L_n(\Omega)$ is then also a Gaussian,
\begin{equation}
\label{eapt2-ft}
L_n(\Omega) = e^{-(\Omega - \omega_n)^2T^2/4P} \, ,
\end{equation}
and the integral in Eq.\ (\ref{cvj-ft}) leads to the Hilbert transform of a
Gaussian function, which does not have a simple analytic
representation.\cite{KING09} Other envelope functions can be easier to handle.
Example is the exponential decay envelope\cite{L95A,ALL96}
$$L_n(t,\tau) = e^{-P|t-\tau|/T}\, .$$ Its Fourier image
$$L_n(\Omega) = \frac{P/T}{(\Omega-\omega_n)^2 + (P/T)^2}$$
has a Lorentzian lineshape and therefore a simple residue
structure which makes the integration of Eq.\ (\ref{cvj-ft})
in the complex plane straightforward. 
The subsequent integrations in 
\begin{equation}
\label{plan}
\int_0^\infty dE_k \int d\tau e^{i\omega_{NFT}\tau} |c_j|^2 \, .
\end{equation}
are also analytical. 
The resulting expressions, however, are awkward and
tedious to analyze. 

In fact, the nature of spectral peaks in the quantum mechanical NFT
spectra can be exposed using an approximation. We replace the pole factor
in Eq.\ (\ref{cvj-ft}) with $-i\pi\delta(\Omega-\epsilon_{vi})$ and ignore
the principal value contribution. 
This is similar to the approximation of temporarily non-overlapping pump
and probe pulses, familiar in the context of ultrafast pump-probe
spectroscopies.\cite{SSD89,SD91A,L95A,WEIHNACHTPEARSON19}
With this approximation, the pump probe amplitude $c_j(v_j;E_k,\tau)$ can be
easily evaluated, and its square is written as 
\begin{eqnarray}
  \left|c_j(v_j;E_k,\tau)\right|^2 & = &
  \sum_{i_1,i_2\ne j}\sum_{vi_1,vi_2}
  \sum_{n_1,n_2,n_3,n_4}
  \mu_{ji_1}\mu_{i_10}(E_k)
  \mu_{ji_2}\mu_{i_20}(E_k)
  \nonumber \\
  &\times&
  \langle v_j | v_{i_1}\rangle\langle v_{i_2} | v_j\rangle
  \langle v_{i_1}|0\rangle\langle 0|v_{i_1}\rangle
  a_{n_1}a_{n_2}a_{n_3}a_{n_4}
  \nonumber \\
  &\times&
  L_{n_1}(\epsilon_{vj}-\epsilon_{vi_1})
  L_{n_2}(\epsilon_{vi_1}-\epsilon_{0} + E_k)
  L_{n_3}(\epsilon_{vj}-\epsilon_{vi_2})
  L_{n_4}(\epsilon_{vi_2}-\epsilon_{0}+E_k)
  \nonumber \\
  &\times& \Big[
  (1+e^{i(\epsilon_{vj}-\epsilon_{vi_1})\tau})
  (1+e^{i(\epsilon_{vi_1}-\epsilon_{0}+E_k)\tau})
  \nonumber \\
  && 
  \,\,\,(1+e^{-i(\epsilon_{vj}-\epsilon_{vi_2})\tau})
  (1+e^{-i(\epsilon_{vi_2}-\epsilon_{0}+E_k)\tau})\Big] \, .
\label{cvj2-ft}
\end{eqnarray}
In this expression,
we expand the electric fields as sums over the odd harmonics, as in
Eq.\ (\ref{eapt1-ft}), and assume that the coefficients $\{a_n\}$ and
the TDMs are real. The dependence
of the probability $|c_j|^2$ on the delay time $\tau$ and the
photoelectron kinetic energy $E_k$  
is now explicit, and the integrals in Eq.\ (\ref{plan}) can be performed
directly. 
Let us define the energy differences between the molecular or ionic states
as $\Delta_{ji} = \epsilon_{vj}-\epsilon_{vi}$ and
$\Delta_{i0} = \epsilon_{vi}-\epsilon_{0}$. Then the NFT spectrum, obtained
after the integrations, reads as
\begin{align}
  \label{nfts-approx}
  \notag I_{\rm ion}(\omega_{\rm NFT}) & = 
  \sum_{j}\sum_{\epsilon_{vj}>A_{\rm ion}} \sum_{i_1,i_2\ne j}\sum_{vi_1,vi_2}
  \sum_{n_1,n_2,n_3,n_4}
  \mu_{ji_1}\mu_{i_10}
  \mu_{ji_2}\mu_{i_20}
  \\
  \notag \times&
  \langle v_j | v_{i_1}\rangle\langle v_{i_2} | v_j\rangle
  \langle v_{i_1}|0\rangle\langle 0|v_{i_1}\rangle
  a_{n_1}a_{n_2}a_{n_3}a_{n_4} \\
  \notag \times&
  \bigg[
  L_{n_1}(\Delta_{ji_1})L_{n_3}(\Delta_{ji_2})
  \frac{1}{2}\sqrt{\frac{2\pi P}{T^2}}e^{-T^2(s_{n_2}-s_{n_4})^2/8P}
  \left(1 + {\rm erf}\left(\frac{T}{\sqrt{2P}}\overline{s_{n_2,n_4}}
  \right)\right) \\
  \notag \times& \Big\{
  2\delta_\tau(\omega_{\rm NFT})
  + \delta_\tau(\omega_{\rm NFT}\pm \Delta_{ji_1})
  + \delta_\tau(\omega_{\rm NFT}\pm \Delta_{ji_2})
  + \delta_\tau(\omega_{\rm NFT}\pm \Delta_{i_1i_2})\Big\}_{1} \\
  \notag + & 
  L_{n_1}(\Delta_{ji_1})L_{n_3}(\Delta_{ji_2})  \\
  \notag \times& \Big\{
  L_{n_2}(\omega_{\rm NFT})L_{n_4}(\omega_{\rm NFT}-\Delta_{i_1i_2})  
  +
  L_{n_2}(\omega_{\rm NFT}+\Delta_{i_1i_2})L_{n_4}(\omega_{\rm NFT})  \\
  \notag & 
  +
  L_{n_2}(\omega_{\rm NFT}-\Delta_{ji_1})
  L_{n_4}(\omega_{\rm NFT}-\Delta_{ji_2})  
  +  
  L_{n_2}(\omega_{\rm NFT}+\Delta_{ji_1})
  L_{n_4}(\omega_{\rm NFT}+\Delta_{ji_2})  
  \Big\}_{2}
  \bigg] \, ; \\
   & \\
  \notag s_{n_2} & =  \omega_{n_2} - \Delta_{i_10} \, ;\\
  \notag s_{n_4} & =  \omega_{n_4} - \Delta_{i_20}\, ; \\
  \notag \overline{s_{n_2,n_4}} & =  \frac{1}{2}(s_{n_2}+s_{n_4}) \, .
  \nonumber 
\end{align}
Here ${\rm erf}(x)$ is the standard error integral.\cite{ABRAMOWITZ70}
The expression holds for any lineshape function $L_n(\Omega)$, and we shall
use Gaussians as in Eq.\ (\ref{eapt2-ft}). 
The NFT spectrum arises as a superposition of ionizations of
the parent molecule into the intermediate ionic states states $i_1$ and
$i_2$ which are then further excited into a dissociative state
$j$. 
There are two groups of NFT spectral peaks in this expression,
corresponding to the two types of 
$\tau$-dependent terms in Eq.\ (\ref{cvj2-ft}). 
The first group --- placed inside the curly brackets $\{\cdots\}_1$
--- includes spectral 
peaks which are located at the energy differences between the intermediate
ionized and the final dissociative states,
$\omega_{\rm NFT} = \pm \Delta_{ji}$. Their
width is controlled by the maximum
delay time in Eq.\ (\ref{ion_w}), and can potentially
be made narrow (delta-function like) by increasing $\tau_{\rm max}$.
These NFT peaks are denoted as $\delta_\tau$. 
They stem from the terms in $|c_j(v_j;E_k,\tau)|^2$ 
with the exponential phase factors independent of $E_k$, such as
$e^{-i\Delta_{ji_1}\tau}$ for example. The positions of these peaks
need not coincide with the multiples of the fundamental harmonic 
frequency $\omega_0$.
The second group
includes terms --- placed inside the curly brackets $\{\cdots\}_2$ ---
which are products of two Gaussian functions $L_n$. Each Gaussian factor peaks
at $\omega_{\rm NFT} = \omega_n \pm \Delta_{ji}$; the 
spectral width is determined
by the reciprocal of the APT temporal width $T$ [see Eq.\ (\ref{eapt2-ft})].
These spectral lines stem from the $\tau$-dependent terms 
with the exponential phase factors
explicitly depending on $E_k$. The positions of these peaks tend to cluster
around the harmonic frequencies $n\omega_0$. 
Note that the widths of different lines in the actual 
NFT spectrum are expected to be sensitive to either $\tau_{\rm max}$ or $T$.

The accuracy of the approximation of Eq.\ (\ref{nfts-approx}) is illustrated
in Fig.\ \ref{nfts_assign} in the main text: Most of the
approximate spectral lines (black) accurately reproduce the positions of the
numerically exact peaks (red). The agreement is worse for the intensities,
which are more sensitive to the coherent two photon effects, 
but the overall spectrum is quite well recognizable.
This
makes the approximation useful in the analysis of the origin of
near harmonic as well as non-harmonic spectral peaks. 
The peaks assigned $\Delta_{fi}$ stem from the terms in $\{\cdots\}_1$;
the peaks assigned $\omega_n$ and $\omega_n \pm \Delta_{fi}$ are due to
the terms in $\{\cdots\}_2$. In particular, 
peaks around $\omega_{\rm NFT}/\omega_0 = 9, 22, 24$ deviating from the
integer harmonic orders are reproduced. 

Although the NFT spectrum in the above equation is already an approximation,
it is still rather bulky and awkward to use in spectral assignments.
In order to simplify the discussion of the
assignment in Sect.\ \ref{assign}, 
we introduce yet another
approximation and assume that the ionizations in Eq.\ (\ref{nfts-approx})
terminate in the same ionic state 
$i_1 = i_2 = i$. 
This is a realistic scenario, at least for the amplitude Set 1 and
$\omega_0 = 1.55$\,eV. We also consider only the ground vibrational
states in all electronic states, drop sums over $vi$,
and rename the final dissociative
states $\epsilon_j > A_{\rm ion}$ as $f$. 
The NFT spectrum is then given by
\begin{eqnarray}
I_{\rm ion}(\omega_{\rm NFT}) & = &
  \sum_{f}\sum_{i\ne f}
  \sum_{n_1,n_2,n_3,n_4}
  \left(\mu_{fi}\mu_{i0}\right)^2
  a_{n_1}a_{n_2}a_{n_3}a_{n_4}
  \nonumber \\
  &\times&
  L_{n_1}(\Delta_{fi})L_{n_3}(\Delta_{fi})\bigg[
  \frac{1}{2}\sqrt{\frac{2\pi P}{T^2}}e^{-T^2(s_{n_2}-s_{n_4})^2/8P}
  \left(1 + {\rm erf}\left(\frac{T}{\sqrt{2P}}\overline{s_{n_2,n_4}}
  \right)\right)
  \nonumber \\
  &\times& \Big\{
  4\delta_\tau(\omega_{\rm NFT})
  + 2\delta_\tau(\omega_{\rm NFT}- \Delta_{fi})
  + 2\delta_\tau(\omega_{\rm NFT}+ \Delta_{fi})
  \Big\}_{1}
  \nonumber \\
  &+& \Big\{
  2L_{n_2}(\omega_{\rm NFT})L_{n_4}(\omega_{\rm NFT})  
  +
  L_{n_2}(\omega_{\rm NFT}-\Delta_{fi})
  L_{n_4}(\omega_{\rm NFT}-\Delta_{fi})
  \nonumber \\
  && 
  L_{n_2}(\omega_{\rm NFT}+\Delta_{fi})
  L_{n_4}(\omega_{\rm NFT}+\Delta_{fi})  
  \Big\}_{2}
  \bigg] \, .
  \label{nfts-approx2}
\end{eqnarray}
The simplified spectrum again consists of two groups of spectral
peaks, those located at the electronic
energy differences $\pm \Delta_{fi}$, and those
located at the unshifted $(\omega_n)$  and shifted,
$(\omega_n \pm \Delta_{fi})$, harmonic frequencies.
There can therefore be a substantial number of spectral peaks located at
non-harmonic frequencies if $\Delta_{fi}$ is not exactly equal to a
multiple of $\omega_0$. Moreover, the Gaussian factors 
indicate that off-resonance excitations within the spectral width of
$L_n$ can be encountered
in an NFT spectrum. This width is of the order of 1\,eV
(i.e. of the order of $\omega_0$) for the typical experimental APTs, so
that deviations from the harmonic frequencies within $\pm \omega_0$ are
not unexpected.

Note that the factors $L_n$, allowing non-resonant excitations within their
spectral widths,
carry as arguments only the energy differences $\Delta_{fi}$
between the ionic states involved in the optical excitation of the ion;
the energy differences between the
initial state of the parent molecule and the ionized state,
$\Delta_{i0} = \epsilon_i - \epsilon_0$, are not involved in the final
expressions. This is an
intrinsic feature of the photochemical pathway of Eq.\ (\ref{chemeq1}):
The mismatch between $\Delta_{i0}$ and the harmonic
frequency $\omega_n > \Delta_{i0}$ can be compensated by the 
kinetic energy $E_k = \omega_n - \Delta_{i0}$ of the
photoelectron ejected in the
first (ionization) step.

The approximate expressions for the NFT spectra, Eqs.\
(\ref{nfts-approx}) and (\ref{nfts-approx2}), 
are further 
discussed in Sect.\ \ref{assign} and used to explain the assignment
scheme for NFT spectra.


\begin{acknowledgments}
  This work was partly is supported by the U.S. Department of Energy, Office of
  Science, Office of Basic Energy Sciences under Contract No.
  DE-AC02-76SF00515.
  S.C. thanks Tomoya Okino and Yasuo Nabekawa for private communications and
  for providing experimental 2D NFT  spectra of CO$_2$ in digital format.
  \end{acknowledgments}


\clearpage
\newpage
\begin{table}
\caption{The set of ab initio 
doublet electronic states of CO$_2^+$ included in the quantum
mechanical calculations of NFT spectra. Shown are
vertical excitation energies $T_e$ (in eV, relative to the equilibrium of the
ground electronic state of CO$_2$) and the
photoionization dipole matrix elements $\mu_{j0}$
with the
ground electronic state of CO$_2$. Calculations are performed
using MRD-CI method. Experimental energies ($T_e$, in eV) and 
assignments, taken from Ref.\ \onlinecite{WRLS88}, are shown
where available. 
} 
\label{abinitio_en}
\begin{ruledtabular}
\begin{tabular}{ccccc}
No. & State  & $T_e$ (calc) & $T_v$ (exp)& $\mu_{j0}$ \\
\hline
1 &$1^2\Pi_g$ & 13.75 & 13.78 & 1.66  \\
2 &$1^2\Pi_u$ & 17.57 & 17.31 & 1.56  \\
3 &$1^2\Sigma_u^+$ & 18.01 & 18.08 & 0.81  \\
4 &$1^2\Sigma_g^+$ & 19.26 & 19.39 & 0.79  \\
5 &$5^2\Pi_u$ & 26.64 & $-$ & 0.52  \\
\hline
6 &$3^2\Sigma_u^-$ & 28.23 & $-$ & 0.00  \\
7 &$3^2\Delta_u$ & 28.27 & $-$ & 0.04  \\
8 &$7^2\Pi_g$ & 29.50 & $-$ & 0.10  \\
9 &$6^2\Pi_u$ & 29.51 & $-$ & 0.08  \\
10 &$4^2\Sigma_u^-$ & 29.98 & $-$ & 0.00  \\
11 &$8^2\Pi_g$ & 30.12 & $-$ & 0.06  \\
12 &$9^2\Pi_g$ & 30.57 & $-$ & 0.07  \\
13 &$5^2\Sigma_u^-$ & 30.59 & $-$ & 0.00  \\
14 &$4^2\Delta_u$ & 30.62 & $-$ & 0.01  \\
15 &$8^2\Pi_u$  & 30.83 & $-$ & 0.18  \\
16 &$10^2\Pi_g$ & 31.06 & $-$ & 0.12  \\
17 &$9^2\Pi_u$ & 31.55 & $-$ & 0.11  \\
18 &$4^2\Sigma_u^+$ & 31.78 & $-$ & 0.13  \\
19 &$5^2\Sigma_u^+$ & 31.79 & $-$ & 0.04  \\
\hline
20 &$n^2\Sigma_g^+$ & 32.10 (32.70) & $-$ & 0.07  \\
21 &$n^2\Pi_u$ & 31.56 (34.06)& $-$ & 0.06  \\
22 &$n^2\Sigma_u^+$ & 32.50 (35.00)& $-$ & 0.02  \\
\end{tabular}
\end{ruledtabular}
\end{table}


\begin{thebibliography}{10}
\bibitem{NOTE-NFTS-02}
{\em {\rm Note that explicit expressions for the NFT spectra can be obtained
  also without approximations using model APTs with an exponential decay
  envelope $L_n(t,\tau)$ [cf. Eq.\ (\ref{eapt1})]. However, the resulting
  expression for $I_{\rm ion}(\omega_{\rm NFT})$ are long and not very useful
  for the analysis.}}
\end{thebibliography}
\end{document}